# Fundamental Limits of Spectrum Sharing for NOMA-Based Cooperative Relaying Under a Peak Interference Constraint

Vaibhav Kumar, *Student Member, IEEE,* Barry Cardiff, *Senior Member, IEEE,* and Mark F. Flanagan, *Senior Member, IEEE*

## Abstract

Non-orthogonal multiple access (NOMA) and spectrum sharing (SS) are two emerging multiple access technologies for efficient spectrum utilization in future wireless communications standards. In this paper, we present the performance analysis of a NOMA-based cooperative relaying system (CRS) in an underlay spectrum sharing scenario, considering a peak interference constraint (PIC), where the peak interference inflicted by the secondary (unlicensed) network on the primary-user (licensed) receiver (PU-Rx) should be less than a predetermined threshold. In the proposed system the relay and the secondary-user receiver (SU-Rx) are equipped with multiple receive antennas and apply selection combining (SC), where the antenna with highest instantaneous signal-to-noise ratio (SNR) is selected, and maximal-ratio combining (MRC), for signal reception. Closed-form expressions are derived for the average achievable rate and outage probabilities for SS-based CRS-NOMA. These results show that for large values of peak interference power, the SS-based CRS-NOMA outperforms the CRS with conventional orthogonal multiple access (OMA) in terms of spectral efficiency. The effect of the interference channel on the system performance is also discussed, and in particular, it is shown that the interference channel between the secondary-user transmitter (SU-Tx) and the PU-Rx has a more severe effect on the average achievable rate as compared

The authors are with School of Electrical and Electronic Engineering, University College Dublin, Ireland (e-mail: vaibhav.kumar@ucdconnect.ie, barry.cardiff@ucd.ie, mark.flanagan@ieee.org).

Part of the content of this paper appeared in the Proc. of the IEEE Global Communications Conference (GLOBECOM'18) - NOMAT5G Workshop, Abu Dhabi, 09-13 Dec. 2018.

This publication has emanated from research conducted with the financial support of Science Foundation Ireland (SFI) and is co-funded under the European Regional Development Fund under Grant Number 13/RC/2077.

to that between the relay and the PU-Rx. A close agreement between the analytical and numerical results confirm the correctness of our rate and outage analysis.

## I. Introduction

With the proliferation of wireless communication technologies, services and applications, one of the major challenges for the successful implementation of future wireless communications standards is that of supporting large-scale heterogeneous data traffic. NOMA has recently been recognized as a promising multiple-access technology for long-term evolution advanced (LTE-A), 5G and beyond-5G wireless networks [1]–[3]. Many variants of NOMA have been suggested in the literature, e.g., power-domain NOMA [4], sparse code multiple access (SCMA), interleave division multiple access (IDMA), low-density spreading (LDS), pattern division multiple access (PDMA) [5] and lattice partition multiple access (LPMA) [6]. Among all these variants, power-domain NOMA has gained widespread popularity because of its implementation-friendly transceiver architecture (we will refer to power-domain NOMA simply as 'NOMA' throughout this paper). It can accommodate several users within the same orthogonal resource block (time, frequency and/or spreading code) via multiplexing them in the power domain at the transmitter and using successive interference cancellation (SIC) at the receiver to remove the messages intended for other users. In the case of NOMA, users with poor channel conditions have a larger share of transmission power, unlike the conventional OMA where more power is allocated to users with strong channel conditions (also known as the water-filling strategy) [3].

Cognitive radio (CR) is another potential solution to the spectrum scarcity problem, as it can enhance the radio spectrum utilization efficiency via SS. The three main SS paradigms include *underlay, overlay* and *interweave* approaches [7]. In an underlay SS system, secondary/unlicensed users (SUs) operate in a frequency band originally owned by a PU such that the interference caused by the SUs on the primary network is lower than a predefined threshold, frequently referred to as the *interference temperature* [8]. Therefore, no limit is directly imposed on the power transmitted from a SU-Tx; it is sufficient that the interference caused at the PU receiver (PU-Rx) is below the threshold. In a fading channel, the secondary network may take advantage of this fact by opportunistically transmitting at high power when the interference channel between SU-Tx and PU-Rx is in a deep fade. A closed-form expression for the secondary channel capacity under a constraint on the (peak or average) interference inflicted on the PU-Rx was presented in [9] for different channel models including additive white Gaussian noise (AWGN), log-normal shadowing, Rayleigh fading and Nakagami-$m$ fading.

The interference from the PU transmitter (PU-Tx) to the SU-Rx, also termed as the primary-to-secondary interference, was not considered in [9] and hence the results derived serve as an upper-bound on the average achievable rate for the secondary network.

A very promising application of NOMA for power-domain multiplexed transmission using a cooperative relaying system (CRS-NOMA) was proposed in [10], where the source was able to deliver two different symbols to the destination in two time slots with the help of a relay. It was shown in [10] that CRS-NOMA outperforms the conventional OMA based decode-and-forward cooperative relaying system in terms of the average achievable rate in the case of Rayleigh fading channels. The performance superiority of CRS-NOMA in Rician fading was shown in [11].

The integration of NOMA into a spectrum sharing system can further enhance the spectral efficiency of a wireless network. Different models of spectrum sharing NOMA networks, including underlay NOMA, overlay NOMA and cognitive NOMA, were discussed in [12], and it was shown that cooperative relaying can improve reception reliability. As such, cooperative spectrum sharing NOMA networks have lower outage probability compared to their non-cooperative counterparts. Interference management, energy efficiency, multi-carrier cognitive NOMA, cognitive MIMO-NOMA, relay selection/user scheduling, physical-layer security and full-duplex cognitive NOMA were recognized as some of the future challenges in the integration of CR and NOMA technologies in [12]. System design guidelines for full-duplex NOMA in CR networks was provided in [13], and it was noted that the performance of half-duplex CR-NOMA systems becomes strictly restricted because of the co-channel interference and the interference constraint imposed by the primary network. A novel secondary NOMA-relay-assisted spectrum sharing scheme was proposed in [14], where first the quality-of-service (QoS) of the PU was guaranteed using MRC and then the sum-rate of the SUs was maximized. In [15], a cooperative NOMA-based spectrum sharing system was proposed, where the primary network shared the spectrum with the secondary network in exchange for a form of cooperation where the secondary transmitter transmitted the primary user's message as well as its own message using NOMA. Using the tools of stochastic geometry, the outage performance of a NOMA-based large scale underlay CR with randomly deployed users was characterized in [16]. The energy efficiency optimization of a multiuser multiple-input single-output (MISO) NOMA system using CR-inspired NOMA subject to an individual QoS constraint for each primary user was discussed in [17]. The outage performance of a two-user decode-and-forward underlay CR-NOMA system was studied in [18]. It was assumed in [18] that the transmission from the relay does not cause any interference to the primary receiver, and also that there is no interference from the primary transmitter to the secondary receivers. The outage performance of a similar system with imperfect channel state information was presented in [19], where the interference from the relay to the primary network and the interference from the primary network to the secondary network was also considered.

In this paper, we analyze the performance of the CRS-NOMA in an underlay SS scenario, where the power transmitted from the SU-Tx and from the relay are constrained by placing a limit on the peak interference power received at the PU-Rx. For clarity of exposition, no other constraint on the transmit power is imposed. While in practice the transmit power from the SU-Tx or the relay is limited by hardware capabilities and other health-related safety considerations, the rates derived in this paper serve as an upper-bound on the capacity of the SS-based CRS-NOMA under a peak interference power constraint. The main contributions of this paper are summarized as follows:

- First, we derive analytical closed-form expressions for the average achievable sum-rate and outage probability for the CRS-NOMA in an underlay SS scenario for the case where all nodes are equipped with a single antenna.
- Next, we derive closed-form expressions for the average achievable sum-rate and outage probability for the SS-based CRS-NOMA in a scenario where the relay and the SU-Rx are equipped with multiple receive antennas. We consider two different diversity combining schemes at both relay and destination for signal reception – SC and MRC. In the case of SC, the closed-form expressions are represented using elementary functions, whereas for MRC, the expressions are represented using Meijer's G-function, the EGBMGF, the Gauss hypergeometric function and some elementary functions.
- We compare the average achievable sum-rate of the SS-based CRS-NOMA with that of the SS-based CRS-OMA and show that the former outperforms the latter for large values of peak interference power at the PU-Rx.
- We also present the asymptotic analysis of the outage probability for SS-based CRS-NOMA.

The rest of this paper is organized as follows: Section II describes the system model for SS-based CRS-NOMA. In Section III, we present a detailed analysis of the average achievable sum-rate and outage probability of the system considering that all nodes are equipped with a single antenna. In Section IV, we present the analysis for the CRS-NOMA where the relay and the SU-Rx are equipped with multiple receive antennas and SC is used for combining the received signals at the respective nodes. Section V deals with the case when both relay and SU-Rx are equipped with multiple receive antennas and MRC is employed for diversity combining. Results

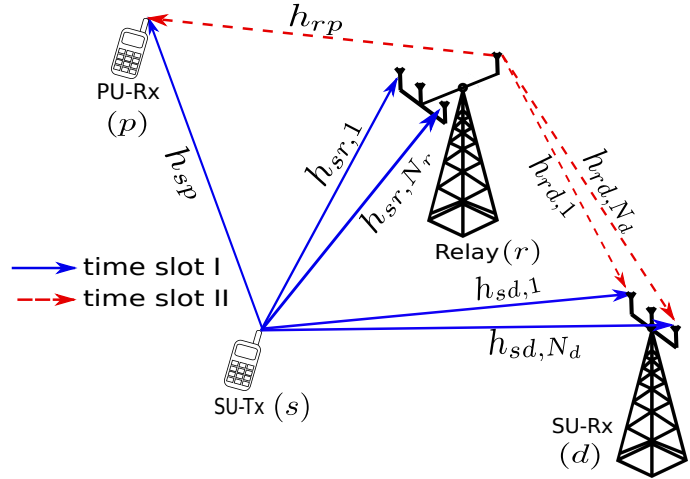

Fig. 1: System model for CRS-NOMA with underlay spectrum sharing.

and discussion are presented in Section VI and finally, conclusions are drawn in Section VII.

## II. System Model

Consider an SS-based CRS-NOMA system as shown in Fig. 1, which consists of a SU-Tx $s$, a relay $r$, a SU-Rx $d$ and a PU-Rx $p$. The SU-Tx $s$ is equipped with a single transmit antenna and the PU-Rx $p$ is equipped with a single receive antenna. The relay $r$ is equipped with $N_r (\geq 1)$ receive antennas and a single transmit antenna, while the SU-Rx $d$ is equipped with $N_d (\geq 1)$ receive antennas. It is assumed that all nodes are operating in the half-duplex mode and all wireless links are independent and Rayleigh distributed. The channel coefficient between the SU-Tx and the $i^{\text{th}}$ receive antenna of the relay $(1 \leq i \leq N_r)$ is denoted by $h_{sr,i}$ and has a mean-square value $\Omega_{sr}$ for all $i$, while that between the SU-Tx and the $j^{\text{th}}$ antenna of the SU-Rx $(1 \leq j \leq N_d)$ is denoted by $h_{sd,j}$ and has a mean-square value $\Omega_{sd}$ for all j. The channel coefficient between the transmit antenna of the relay and the $j^{\text{th}}$ antenna of the SU-Rx is denoted by $h_{rd,j}$ and has a mean-square value $\Omega_{rd}$ for all j. Moreover, the channel coefficient between the SU-Tx and the PU-Rx is denoted by $h_{sp}$ and has a mean-square value $\Omega_{sp}$, while that between the relay and the PU-Rx is denoted by $h_{rp}$ and has a mean-square value $\Omega_{rp}$. Furthermore, it is assumed that the channels between the SU-Tx and the SU-Rx are on average weaker that those between the SU-Tx and the relay, i.e., $\Omega_{sd} < \Omega_{sr}$.

We assume that perfect knowledge of $h_{sp}$ and $h_{rp}$ are available at the SU-Tx and the relay, respectively; this helps in determining the transmit power so that the received interference power at the PU-Rx never exceeds the tolerance limit Q. Also, the knowledge of $\Omega_{sr}$ and $\Omega_{sd}$ is assumed to be available at the SU-Tx. Moreover, we assume that perfect knowledge of $\{h_{sr,i}\}$ is available at the relay, and perfect knowledge of $\{h_{sd,j}\}$ and $\{h_{rd,k}\}$ is available at the SU-Rx. It is important to note that these CSI requirements are similar to those assumed in [9] and [10].

In the CRS-NOMA scheme with spectrum sharing, the SU-Tx broadcasts $\sqrt{a_1 P_s(h_{sp})}s_1 + \sqrt{a_2 P_s(h_{sp})}s_2$ to both relay and SU-Rx, where $s_1$ and $s_2$ are the data-bearing constellation symbols which are multiplexed in the power domain ($\mathbb{E}\{|s_i|^2\} = 1$ for $i = 1, 2$). $P_s(h_{sp})$ is the power transmitted from the SU-Tx and in general is a mapping from the fading coefficient $h_{sp}$ to the set of non-negative real numbers $\mathbb{R}_+$ such that the instantaneous interference at the PU-Rx does not exceed a predetermined value (interference temperature). Moreover, $a_1$ and $a_2$ are power weighting coefficients satisfying the constraints $a_1 + a_2 = 1$ and $a_1 > a_2$. After receiving the signal from the SU-Tx, the SU-Rx decodes symbol $s_1$ treating the interference from $s_2$ as additional noise, while the relay first decodes symbol $s_1$ and then applies SIC to decode $s_2$. In the second time slot, the SU-Tx remains silent and only the relay transmits its estimate of symbol $s_2$, denoted as $\hat{s}_2$, to the SU-Rx with power $P_r(h_{rp})$ which in general is a mapping from the fading coefficient $h_{rp}$ to $\mathbb{R}_+$ such that the instantaneous interference at the primary receiver does not exceed the predetermined threshold. In this manner, two different symbols are delivered to the secondary receiver in two time slots.

In contrast to this, in the conventional OMA with underlay spectrum sharing scheme, the SU-Tx broadcasts symbol $s_1$ with power $P_s(h_{sp})$ in the first time slot and the relay retransmits its estimate of symbol $s_1$, denoted by $\hat{s}_1$, to the SU-Rx with power $P_r(h_{rp})$ in the second time slot. The SU-Rx then combines both copies of symbol $s_1$ and in this manner only a single symbol is delivered to the SU-Rx in two time slots.

## III. Performance Analysis: Single Antenna Case

In this section we present the average achievable sum-rate and outage probability analysis for the SS-based CRS-NOMA under peak interference constraint for the case when $N_r = N_d = 1$. Note that while the more general case ($N_r \geq 1, N_d \geq 1$) will be considered later, for this special case the derived expressions for the average achievable rate and outage probability have a particularly simple form which is not trivial to deduce from the solution to the general case; thus it is worthwhile to consider the single antenna case separately.

Here we denote the channel coefficients for s-r, s-d, r-d, s-p and r-p links by $h_{sr}$, $h_{sd}$, $h_{rd}$, $h_{sp}$ and $h_{rp}$ respectively. The signal received at the relay (resp. SU-Rx and PU-Rx) in the first

time slot is given by

$$y_{s\ell} = h_{s\ell}\left(\sqrt{a_1 P_s(h_{sp})}s_1 + \sqrt{a_2 P_s(h_{sp})}s_2\right) + n_{s\ell},$$

where $\ell = r$ (resp. $\ell = d$ and $\ell = p$) and, $n_{s\ell}$ is complex additive white Gaussian noise (AWGN) with zero mean and unit variance. The received instantaneous signal-to-interference-plus-noise ratio (SINR) at the relay for decoding symbol $s_1$ and the received instantaneous SNR for decoding symbol $s_2$ (assuming symbol $s_1$ is decoded correctly) are, respectively,

$$\gamma_{sr}^{(1)} = \frac{\lambda_{sr}a_1 P_s(h_{sp})}{\lambda_{sr}a_2 P_s(h_{sp}) + 1}, \qquad \gamma_{sr}^{(2)} = \lambda_{sr}a_2 P_s(h_{sp}),$$

where $\lambda_{sr} \triangleq |h_{sr}|^2$. Similarly, the received instantaneous SINR at the SU-Rx for decoding symbol $s_1$ is given by

$$\gamma_{sd} = \frac{\lambda_{sd}a_1 P_s(h_{sp})}{\lambda_{sd}a_2 P_s(h_{sp}) + 1},$$

where $\lambda_{sd} \triangleq |h_{sd}|^2$. In the next time slot, the relay transmits its estimate of symbol $s_2$, denoted by $\hat{s}_2$, to the SU-Rx with power $P_r(h_{rp})$. The signal received at the SU-Rx (resp. PU-Rx) in the second time slot is given by

$$y_{r\sigma} = h_{r\sigma}\sqrt{P_r(h_{rp})}\hat{s}_2 + n_{r\sigma},$$

where $\sigma = d$ (resp. $\sigma = p$) and $n_{r\sigma}$ is complex AWGN with zero mean and unit variance. The received SNR at the SU-Rx for decoding symbol $s_2$ is given by

$$\gamma_{rd} = \lambda_{rd}P_r(h_{rp}),$$

where $\lambda_{rd} \triangleq |h_{rd}|^2$. Since the symbol $s_1$ should be decoded correctly at the SU-Rx as well as at the relay for SIC, while satisfying the interference constraint at the PU-Rx, the average achievable rate for symbol $s_1$ is given by (c.f. [10, Eqn. (8)], [9, Eqn. (1), Eqn. (22)])

$$\bar{C}_{s_1} = \max_{P_s(h_{sp}) \geq 0} \int_{|h_{sp}|} \int_{|h_{sr}|} \int_{|h_{sd}|} 0.5 \log_2\left(1 + \min\left\{\gamma_{sr}^{(1)}, \gamma_{sd}\right\}\right) g_1(|h_{sp}|)g_2(|h_{sr}|)g_3(|h_{sd}|) \, d|h_{sp}|$$

$$\times \, d|h_{sr}| \, d|h_{sd}|, \quad (1)$$

$$\text{s.t.} \quad \lambda_{sp}P_s(h_{sp}) \leq Q, \quad (2)$$

where the maximization in (1) is performed over all possible transmit power mappings $P_s(h_{sp}) > 0$ and $Q$ is the peak interference power that the PU-Rx can tolerate from the secondary network; and $g_1(|h_{sp}|)$, $g_2(|h_{sr}|)$ and $g_3(|h_{sd}|)$ denote the probability density functions (PDFs) of $|h_{sp}|, |h_{sr}|$ and $|h_{sd}|$ respectively. Assuming no other limitation on the power transmitted from the SU-Tx, the optimal transmit power $P_s^*(h_{sp})$ which maximizes the achievable rate is given by $Q/\lambda_{sp}$. Hence, the average achievable rate for symbol $s_1$ is given by

$$\bar{C}_{s_1} = 0.5 \int_{|h_{sp}|} \int_{|h_{sr}|} \int_{|h_{sd}|} \log_2 \left( 1 + \frac{\frac{\min\{\lambda_{sr},\lambda_{sd}\}}{\lambda_{sp}} Q a_1}{\frac{\min\{\lambda_{sr},\lambda_{sd}\}}{\lambda_{sp}} Q a_2 + 1} \right) g_1(|h_{sp}|) g_2(|h_{sr}|) g_3(|h_{sd}|) \, d|h_{sp}|$$

$$\times \, d|h_{sr}| \, d|h_{sd}|$$

$$= \frac{1}{2} \left[ \int_0^\infty \log_2(1+Qx) f_X(x) dx - \int_0^\infty \log_2(1+Qa_2 x) f_X(x) dx \right], \tag{3}$$

where $X \triangleq \min\{\lambda_{sr}, \lambda_{sd}\}/\lambda_{sp}$ and $f_{\mathcal{W}}(\cdot)$ denotes the PDF of the random variable $\mathcal{W}$.

**Theorem 1.** *A closed-form expression for the average achievable rate for the symbol $s_1$ is given by*

$$\bar{C}_{s_1} = 0.5 \left[ \frac{Q \log_2 \left( \frac{Q}{\phi \Omega_{sp}} \right)}{Q - \phi \Omega_{sp}} - \frac{a_2 Q \log_2 \left( \frac{a_2 Q}{\phi \Omega_{sp}} \right)}{a_2 Q - \phi \Omega_{sp}} \right], \tag{4}$$

*where $\phi \triangleq (1/\Omega_{sr}) + (1/\Omega_{sd})$.*

*Proof*: See Appendix A.

Similarly, the average achievable rate for symbol $s_2$ is given by (c.f. [10, Eqn. (9)], [9, Eqn. (1), Eqn. (22)])

$$\bar{C}_{s_2} = \max_{\substack{P_s(h_{sp}) \geq 0 \\ P_r(h_{rp}) \geq 0}} \int_{|h_{sp}|} \int_{|h_{rp}|} \int_{|h_{sr}|} \int_{|h_{rd}|} 0.5 \log_2 \left( 1 + \min \left\{ \gamma_{sr}^{(2)}, \gamma_{rd} \right\} \right) g_1(|h_{sp}|) \, g_2(|h_{sr}|) \, g_4(|h_{rp}|) \, g_5(|h_{rd}|)$$

$$\times \, d|h_{sp}| \, d|h_{sr}| \, d|h_{rp}| \, d|h_{rd}|, \tag{5}$$

s.t.   $\lambda_{sp} P_s(h_{sp}) \leq Q,$ \hfill (6)

$\lambda_{rp} P_r(h_{rp}) \leq Q,$ \hfill (7)

where $g_4(|h_{rp}|)$ and $g_5(|h_{rd}|)$ denote the PDFs of $|h_{rp}|$ and $|h_{rd}|$, respectively. Assuming no other limitation on the power transmitted from the relay, the optimal transmit power $P_r^*(h_{rp})$ which maximizes the achievable rate is given by $Q/\lambda_{rp}$. Therefore, the average achievable rate for symbol $s_2$ is given by

$$\bar{C}_{s_2} = \int_{|h_{sp}|} \int_{|h_{rp}|} \int_{|h_{sr}|} \int_{|h_{rd}|} 0.5 \log_2 \left( 1 + \min \left\{ \frac{\lambda_{sr} a_2}{\lambda_{sp}}, \frac{\lambda_{rd}}{\lambda_{rp}} \right\} Q \right) g_1(|h_{sp}|) \, g_2(|h_{sr}|) \, g_4(|h_{rp}|) \, g_5(|h_{rd}|)$$

$$\times \, d|h_{sp}| \, d|h_{sr}| \, d|h_{rp}| \, d|h_{rd}|$$

$$= \frac{1}{2} \int_0^\infty \log_2(1+Qx) f_Y(x) \, dx = \frac{0.5Q}{\ln(2)} \int_0^\infty \frac{1 - F_Y(x)}{1 + Qx} \, dx, \tag{8}$$

where $Y \triangleq \min\{\lambda_{sr} a_2/\lambda_{sp}, \lambda_{rd}/\lambda_{rp}\}$ and $F_{\mathcal{W}}(\cdot)$ denotes the cumulative distribution function (CDF) of the random variable $\mathcal{W}$.

**Theorem 2.** *A closed-form expression for the average achievable rate for the symbol* $s_2$ *is given by*

$$\bar{C}_{s_2} = \frac{0.5\, a_2\, \Omega_{rd}\, \Omega_{sr}\, Q \left[ \Omega_{rp} \Omega_{sp} \log_2 \left( \frac{a_2 \Omega_{rp} \Omega_{sr}}{\Omega_{rd} \Omega_{sp}} \right) + a_2 \Omega_{rp} \Omega_{sr} Q \log_2 \left( \frac{\Omega_{rd} Q}{\Omega_{rp}} \right) - \Omega_{rd} \Omega_{sp} Q \log_2 \left( \frac{a_2 \Omega_{sr} Q}{\Omega_{sp}} \right) \right]}{(\Omega_{rd} \Omega_{sp} - a_2 \Omega_{rp} \Omega_{sr})(\Omega_{rd} Q - \Omega_{rp})(\Omega_{sp} - a_2 \Omega_{sr} Q)}.$$

$$(9)$$

*Proof*: See Appendix B.

Using (4) and (9), the average achievable sum-rate for the CRS-NOMA system is given as

$$\bar{C}_{sum} = \bar{C}_{s_1} + \bar{C}_{s_2}. \tag{10}$$

In theory the system with single antenna (at the relay and SU-Rx) are special case when the relay and SU-Rx are equipped with multiple receive antennas, however, the analysis leads to simple closed-form expressions for the average achievable sum-rate and outage probability, and these expressions are not trivial to obtain from the generalized case of multiple antennas.

For the case of OMA, the signal received at the relay (resp. SU-Rx and PU-Rx) in the first time slot is given by

$$y_{s\ell,\text{OMA}} = h_{s\ell} \sqrt{P_s(h_{sp})} s_1 + n_{s\ell},$$

where $\ell = r$ (resp. $\ell = d$ and $\ell = p$). In the second time slot, the relay transmits its estimate of $s_1$, denoted by $\hat{s}_1$, to the SU-Rx. The signal received at the SU-Rx (resp. PU-Rx) in the second time slot is given by

$$y_{r\sigma,\text{OMA}} = h_{r\sigma} \sqrt{P_r(h_{rp})} \hat{s}_1 + n_{r\sigma},$$

where $\sigma = d$ (resp. $\sigma = p$). Following the same peak interference constraint as in the case of NOMA, the average achievable rate for the SS-based CRS-OMA is given by

$$\bar{C}_{\text{OMA}} = 0.5 \mathbb{E}_Z \left[ \log_2(1 + QZ) \right], \tag{11}$$

where[1] $Z \triangleq \min \left\{ \frac{\lambda_{sr}}{\lambda_{sp}}, \frac{\lambda_{sd}}{\lambda_{sp}} + \frac{\lambda_{rd}}{\lambda_{rp}} \right\}$ and $\mathbb{E}_{\mathcal{W}}[\cdot]$ denotes the expectation with respect to the random variable $\mathcal{W}$.

---

[1]Here we assume that the SU-Rx applies MRC on the two copies of $s_1$.

*Outage probability for CRS-NOMA:* We define $\mathcal{O}_1$ as the outage event for symbol $s_1$, i.e., the event when either the relay or the SU-Rx fails to decode $s_1$ successfully. Hence the outage probability for symbol $s_1$ is given by

$$\Pr(\mathcal{O}_1) = \Pr(C_{s_1} < R_1) = \Pr\left[0.5\log_2\left(1 + \frac{a_1 QX}{a_2 QX + 1}\right) < R_1\right]$$

$$= \Pr(X < \Theta_1) = \int_0^{\Theta_1} f_X(x)\,dx = \int_0^{\Theta_1}\frac{\phi\Omega_{sp}\,dx}{(1 + \phi\Omega_{sp}x)^2} = \frac{\phi\Omega_{sp}\Theta_1}{1 + \phi\Omega_{sp}\Theta_1}, \quad (12)$$

where $C_{s_1}$ is the instantaneous achievable rate for symbol $s_1$, $R_1$ is the target data rate for symbol $s_1$, $\epsilon_1 = 2^{2R_1} - 1$ and $\Theta_1 = \frac{\epsilon_1}{Q(a_1 - \epsilon_1 a_2)}$. The integration above is solved using (33) and [20, Eqn. (3.194-1), p. 315]. The system design must ensure that $a_1 > \epsilon_1 a_2$, otherwise the outage probability for symbol $s_1$ will always be 1 as noted in [21]. Next, we define $\mathcal{O}_2$ as the outage event for symbol $s_2$. This outage event can be decomposed as the union of the following disjoint events: (i) symbol $s_1$ cannot be successfully decoded at the relay; (ii) symbol $s_1$ is successfully decoded at the relay, but symbol $s_2$ cannot be successfully decoded at the relay; and (iii) both symbols are successfully decoded at the relay, but symbol $s_2$ cannot be successfully decoded at the SU-Rx. Therefore, the outage probability for the symbol $s_2$ may be expressed as

$$\Pr(\mathcal{O}_2) = \begin{cases} \Pr\left(\frac{\lambda_{sr}}{\lambda_{sp}} < \Theta_1\right) + \Pr\left(\frac{\lambda_{sr}}{\lambda_{sp}} \geq \Theta_1, \frac{\lambda_{sr}}{\lambda_{sp}} < \Theta_2\right) + \Pr\left(\frac{\lambda_{sr}}{\lambda_{sp}} \geq \Theta_2, \frac{\lambda_{rd}}{\lambda_{rp}} < \frac{\epsilon_2}{Q}\right); & \text{if } \Theta_1 < \Theta_2 \\ \Pr\left(\frac{\lambda_{sr}}{\lambda_{sp}} < \Theta_1\right) + \Pr\left(\frac{\lambda_{sr}}{\lambda_{sp}} \geq \Theta_1, \frac{\lambda_{rd}}{\lambda_{rp}} < \frac{\epsilon_2}{Q}\right); & \text{otherwise} \end{cases}$$

$$= F_{\frac{\lambda_{sr}}{\lambda_{sp}}}(\Theta) + F_{\frac{\lambda_{rd}}{\lambda_{rp}}}\left(\frac{\epsilon_2}{Q}\right) - F_{\frac{\lambda_{sr}}{\lambda_{sp}}}(\Theta)F_{\frac{\lambda_{rd}}{\lambda_{rp}}}\left(\frac{\epsilon_2}{Q}\right), \quad (13)$$

where $R_2$ is the target data rate for symbol $s_2$, $\epsilon_2 = 2^{2R_2} - 1$, $\Theta_2 = \frac{\epsilon_2}{a_2 Q}$ and $\Theta = \max(\Theta_1, \Theta_2)$. Using (35), the closed-form expression for the outage probability of symbol $s_2$ is given by

$$\Pr(\mathcal{O}_2) = \frac{\Omega_{sp}\Theta}{\Omega_{sr} + \Omega_{sp}\Theta} + \frac{\epsilon_2\Omega_{rp}}{Q\Omega_{rd} + \epsilon_2\Omega_{rp}} - \frac{\Omega_{sp}\Omega_{rp}\epsilon_2\Theta}{(\Omega_{sr} + \Omega_{sp}\Theta)(\Omega_{rd}Q + \Omega_{rp}\epsilon_2)}. \quad (14)$$

On the other hand, for the SS-based CRS-OMA system, the outage probability is given by

$$\Pr(\mathcal{O}_{OMA}) = \Pr(Z < \epsilon_1).$$

*Asymptotic behavior of CRS-NOMA*

Using (12), for large values of $Q$ we have

$$\Pr(\mathcal{O}_1) = \frac{\phi\Omega_{sp}\epsilon_1}{Q(a_1 - \epsilon_1 a_2) + \phi\Omega_{sp}\epsilon_1} = \frac{1}{1 + \beta Q} = \frac{1}{\beta Q}\sum_{k=0}^{\infty}(-1)^k\left(\frac{1}{\beta Q}\right)^k = \frac{1}{\beta Q} + \mathbf{O}\left(Q^{-2}\right),$$

where $\beta \triangleq (a_1 - \epsilon_1 a_2)/(\phi \Omega_{sp} \epsilon_1)$ and $\mathbf{O}(\cdot)$ is the Landau symbol. The above relation holds good for $\beta Q > 1$. Therefore, it is clear that for the single-antenna case, the outage probability for $s_1$ decays as $Q^{-1}$ for large $Q$. Similarly, for large values of $Q$

$$F_{\frac{\lambda_{sr}}{\lambda_{sp}}}(\Theta) = \frac{\Omega_{sp}\Theta}{\Omega_{sr} + \Omega_{sp}\Theta} = \frac{\Omega_{sp}\beta^\dagger}{\Omega_{sr}Q} \sum_{k=0}^{\infty} (-1)^k \left(\frac{\Omega_{sp}\beta^\dagger}{\Omega_{sr}Q}\right)^k = \frac{\Omega_{sp}\beta^\dagger}{\Omega_{sr}Q} + \mathbf{O}\left(Q^{-2}\right),$$

where $\beta^\dagger \triangleq \max\left\{\frac{\epsilon_1}{a_1 - \epsilon_1 a_2}, \frac{\epsilon_2}{a_2}\right\}$. Hence, $F_{\frac{\lambda_{sr}}{\lambda_{sp}}}(\Theta)$ decays as $Q^{-1}$ for large $Q$ values. Analogously, it can be shown that $F_{\frac{\lambda_{rd}}{\lambda_{rp}}}\left(\frac{\epsilon_2}{Q}\right)$ decays as $Q^{-1}$ and $F_{\frac{\lambda_{sr}}{\lambda_{sp}}}(\Theta)F_{\frac{\lambda_{rd}}{\lambda_{rp}}}\left(\frac{\epsilon_2}{Q}\right)$ decays as $Q^{-2}$. Therefore, using (13), it is clear that the outage probability for $s_2$ decays as $Q^{-1}$ for large values of $Q$.

## IV. Performance Analysis: Selection Combining

In this section, we generalize the results obtained in the previous section for the case when $N_r \geq 1$, $N_d \geq 1$ and selection combining (SC), i.e., selection of the antenna with highest instantaneous SNR, is used for reception at both relay and SU-Rx. In the first time-slot, the received instantaneous SINR at the relay for decoding symbol $s_1$ and the instantaneous SNR for decoding symbol $s_2$ (assuming the symbol $s_1$ is decoded correctly) are, respectively,

$$\gamma_{sr,SC}^{(1)} = \frac{\delta_{sr}a_1 P_s(h_{sp})}{\delta_{sr}a_2 P_s(h_{sp}) + 1}, \quad \gamma_{sr,SC}^{(2)} = \delta_{sr}a_2 P_s(h_{sp}),$$

where $i^* \triangleq \operatorname{argmax}_{1 \leq i \leq N_r} |h_{sr,i}|$ and $\delta_{sr} \triangleq |h_{sr,i^*}|^2$. Similarly, the received instantaneous SINR at the SU-Rx for decoding symbol $s_1$ in the first time slot is given by

$$\gamma_{sd,SC} = \frac{\delta_{sd}a_1 P_s(h_{sp})}{\delta_{sd}a_2 P_s(h_{sp}) + 1},$$

where $j^* \triangleq \operatorname{argmax}_{1 \leq j \leq N_d} |h_{sd,j}|$ and $\delta_{sd} \triangleq |h_{sd,j^*}|^2$. The received SNR at the SU-Rx for decoding symbol $s_2$ in the second time slot is given by

$$\gamma_{rd,SC} = \delta_{rd}P_r(h_{rp}),$$

where $k^* \triangleq \operatorname{argmax}_{1 \leq k \leq N_d} |h_{rd,k}|$ and $\delta_{rd} \triangleq |h_{rd,k^*}|^2$. Following similar arguments as in the previous section, the average achievable rate for symbol $s_1$ using SC is given by

$$\bar{C}_{s_1,SC} = 0.5 \int_0^\infty \log_2(1 + Qx)f_{\mathcal{X}}(x)\,dx - 0.5 \int_0^\infty \log_2(1 + Qa_2 x)f_{\mathcal{X}}(x)\,dx, \qquad (15)$$

where $\mathcal{X} \triangleq \min\{\delta_{sr}, \delta_{sd}\}/\lambda_{sp}$.

**Theorem 3.** *A closed-form expression for the average achievable rate for symbol* $s_1$ *using SC is given by*

$$\bar{C}_{s_1,SC} = 0.5 \sum_{k=1}^{N_r} \sum_{j=1}^{N_d} (-1)^{k+j} \binom{N_r}{k} \binom{N_d}{j} \left[\frac{Q \log_2\left(\frac{Q}{\xi_{k,j}\Omega_{sp}}\right)}{Q - \xi_{k,j}\Omega_{sp}} - \frac{a_2 Q \log_2\left(\frac{a_2 Q}{\xi_{k,j}\Omega_{sp}}\right)}{a_2 Q - \xi_{k,j}\Omega_{sp}}\right], \qquad (16)$$

*where* $\xi_{k,j} \triangleq (k/\Omega_{sr}) + (j/\Omega_{sd})$.

*Proof*: See Appendix C.

Similarly, the average achievable rate for symbol $s_2$ using SC is given by

$$\bar{C}_{s_2,SC} = 0.5 \int_0^\infty \log_2(1 + Qx)\, f_{\mathcal{Y}}(x)\, dx, \tag{17}$$

where $\mathcal{Y} \triangleq \min\{\delta_{sr} a_2 / \lambda_{sp}, \delta_{rd} / \lambda_{rp}\}$.

**Theorem 4.** *A closed-form expression for the average achievable rate for symbol $s_2$ using SC is given by*

$$\bar{C}_{s_2,SC} = 0.5Q \left[ \sum_{k=1}^{N_r} (-1)^{k-1} \binom{N_r}{k} \frac{a_2 \Omega_{sr} \log_2\left(\frac{a_2 \Omega_{sr} Q}{k \Omega_{sp}}\right)}{a_2 \Omega_{sr} Q - k \Omega_{sp}} + \sum_{j=1}^{N_d} (-1)^{j-1} \binom{N_d}{j} \frac{\Omega_{rd} \log_2\left(\frac{\Omega_{rd} Q}{j \Omega_{rp}}\right)}{\Omega_{rd} Q - j \Omega_{rp}} \right.$$

$$+ \sum_{k=1}^{N_r} \sum_{j=1}^{N_d} (-1)^{k+j} \binom{N_r}{k} \binom{N_d}{j} \left\{ \frac{k \Omega_{rd}^2 \Omega_{sp} \log_2\left(\frac{j \Omega_{rp}}{Q \Omega_{rd}}\right)}{(k \Omega_{rd} \Omega_{sp} - j a_2 \Omega_{rp} \Omega_{sr})(Q \Omega_{rd} - j \Omega_{rp})} \right.$$

$$\left. \left. + \frac{j a_2^2 \Omega_{rp} \Omega_{sr}^2 \log_2\left(\frac{a_2 Q \Omega_{sr}}{k \Omega_{sp}}\right)}{(k \Omega_{rd} \Omega_{sp} - j a_2 \Omega_{rp} \Omega_{sr})(a_2 Q \Omega_{sr} - k \Omega_{sp})} \right\} \right]. \tag{18}$$

*Proof*: See Appendix D.

Using (16) and (18), the average achievable sum-rate for the CRS-NOMA system using SC is given by

$$\bar{C}_{sum,SC} = \bar{C}_{s_1,SC} + \bar{C}_{s_2,SC}. \tag{19}$$

With extensive algebraic manipulations, it can be shown that for $N_r = N_d = 1$, (16) and (18) reduces to (4) and (9), respectively. For the case of SS-based CRS-OMA with SC, the average achievable rate is given as

$$\bar{C}_{OMA,SC} = 0.5 \mathbb{E}_{\mathcal{Z}} \left[ \log_2(1 + Q\mathcal{Z}) \right], \tag{20}$$

where $\mathcal{Z} \triangleq \min\{\frac{\delta_{sr}}{\lambda_{sp}}, \frac{\delta_{sd}}{\lambda_{sp}} + \frac{\delta_{rd}}{\lambda_{rp}}\}$.

*Outage probability for CRS-NOMA with SC:* Similar to the previous section, we define $\mathcal{O}_1$ as the outage event for symbol $s_1$. Hence the outage probability for symbol $s_1$ using SC is given by

$$\Pr(\mathcal{O}_1) = \Pr\left(C_{s_1,SC} < R_1\right) = \Pr(\mathcal{X} < \Theta_1) = \int_0^{\Theta_1} f_{\mathcal{X}}(x) dx$$

$$= \sum_{k=1}^{N_r} \sum_{j=1}^{N_d} (-1)^{k+j} \binom{N_d}{j} \binom{N_r}{k} \frac{\xi_{k,j} \Omega_{sp} \Theta_1}{1 + \xi_{k,j} \Omega_{sp} \Theta_1}, \tag{21}$$

where $C_{s_1,SC}$ is the instantaneous achievable rate for symbol $s_1$ in the SS-based CRS-NOMA using SC. The integration above is solved using (40) and [20, Eqn. (3.194-1), p. 315]. Next, we define $\mathcal{O}_2$ as the outage event for symbol $s_2$ using SC, similar to the previous section. Hence, the outage probability for symbol $s_2$ is given by

$$\Pr(\mathcal{O}_2) = F_{\frac{\delta_{sr}}{\lambda_{sp}}}(\Theta) + F_{\frac{\delta_{rd}}{\lambda_{rp}}}\left(\frac{\epsilon_2}{Q}\right) - F_{\frac{\delta_{sr}}{\lambda_{sp}}}(\Theta) F_{\frac{\delta_{rd}}{\lambda_{rp}}}\left(\frac{\epsilon_2}{Q}\right). \tag{22}$$

Using (22) and (42), the closed-form expression for the outage probability of symbol $s_2$ in SS-based CRS-NOMA using SC is given by

$$\Pr(\mathcal{O}_2) = \sum_{k=1}^{N_r} \binom{N_r}{k} \frac{(-1)^{k-1} k \Omega_{sp}\Theta}{\Omega_{sr} + k\Omega_{sp}\Theta} + \sum_{j=1}^{N_d} \binom{N_d}{j} \frac{(-1)^{j-1} j \epsilon_2 \Omega_{rp}}{Q\Omega_{rd} + j\epsilon_2 \Omega_{rp}}$$

$$- \sum_{k=1}^{N_r} \sum_{j=1}^{N_d} \binom{N_r}{k}\binom{N_d}{j} \frac{(-1)^{k+j} kj\Omega_{sp}\Omega_{rp}\epsilon_2\Theta}{(\Omega_{sr} + k\Omega_{sp}\Theta)(\Omega_{rd}Q + j\Omega_{rp}\epsilon_2)}. \tag{23}$$

On the other hand, the outage probability for the SS-based CRS-OMA with SC is given by

$$\Pr(\mathcal{O}_{OMA}) = \Pr(\mathcal{Z} < \epsilon_1).$$

*Asymptotic behavior of CRS-NOMA with SC*

Using (21), for large values of $Q$,

$$\Pr(\mathcal{O}_1) = \sum_{k=1}^{N_r} \sum_{j=1}^{N_d} \sum_{n=0}^{\infty} (-1)^{k+j+n} \binom{N_d}{j}\binom{N_r}{k} \xi_{k,j}^{n+1} \left[\frac{\Omega_{sp}\epsilon_1}{(a_1 - \epsilon_1 a_2)Q}\right]^{n+1}.$$

Using [20, Eqn. (0.154-3), p. 4], we have

$$\Pr(\mathcal{O}_1) = \sum_{k=1}^{N_r} \sum_{j=1}^{N_d} \sum_{n=\min\{N_r,N_d\}-1}^{\infty} (-1)^{k+j+n} \binom{N_d}{j}\binom{N_r}{k} \xi_{k,j}^{n+1} \left[\frac{\Omega_{sp}\epsilon_1}{(a_1 - \epsilon_1 a_2)Q}\right]^{n+1} \propto Q^{-\min\{N_r,N_d\}}.$$

Therefore, it is clear that for the case of SS-based CRS-NOMA, the outage probability for $s_1$ decays as $Q^{-\min\{N_r,N_d\}}$ for large $Q$ values. Using a similar identity, it can be shown that $F_{\delta_{sr}/\lambda_{sp}}(\Theta)$ decays as $Q^{-N_r}$, $F_{\delta_{rd}/\lambda_{rp}}(\epsilon_2/Q)$ decays as $Q^{-N_d}$ and $F_{\delta_{sr}/\lambda_{sp}}(\Theta)F_{\delta_{rd}/\lambda_{rp}}(\epsilon_2/Q)$ decays as $Q^{-(N_r+N_d)}$ for large values of $Q$. Therefore, using (22), it follows that for the case of SS-based CRS-NOMA, the outage probability for $s_2$ decays as $Q^{-\min\{N_r,N_d\}}$ for large values of $Q$.

## V. Performance Analysis: Maximal Ratio Combining

In this section, we analyze the scenario when $N_r \geq 1$, $N_d \geq 1$ and MRC is used for combining the signals at the relay and the SU-Rx. The received instantaneous SINR at the relay for decoding symbol $s_1$ and the instantaneous SNR for decoding symbol $s_2$ (assuming the symbol $s_1$ is decoded correctly) in the first time slot are, respectively,

$$\gamma_{sr,MRC}^{(1)} = \frac{\eta_{sr}a_1 P_s(h_{sp})}{\eta_{sr}a_2 P_s(h_{sp}) + 1}, \qquad \gamma_{sr,MRC}^{(2)} = \eta_{sr}a_2 P_s(h_{sp}),$$

where $\eta_{sr} \triangleq \sum_{i=1}^{N_r} |h_{sr,i}|^2$. Similarly, the received instantaneous SINR at the SU-Rx for decoding symbol $s_1$ in the first time slot is given by

$$\gamma_{sd,MRC} = \frac{\eta_{sd}a_1 P_s(h_{sp})}{\eta_{sd}a_2 P_s(h_{sp}) + 1},$$

where $\eta_{sd} \triangleq \sum_{j=1}^{N_d} |h_{sd,j}|^2$. The received SNR at the SU-Rx for decoding symbol $s_2$ in the second time slot is given by

$$\gamma_{rd,MRC} = \eta_{rd} P_r(h_{rp}),$$

where $\eta_{rd} \triangleq \sum_{k=1}^{N_d} |h_{rd,k}|^2$. The average achievable rate for symbol $s_1$ using MRC is given by

$$\bar{C}_{s_1,MRC} = 0.5 \int_0^\infty \log_2(1 + Qx) f_{\mathcal{X}}(x)\, dx - 0.5 \int_0^\infty \log_2(1 + Qa_2 x) f_{\mathcal{X}}(x)\, dx, \qquad (24)$$

where $\mathcal{X} \triangleq \min\{\eta_{sr}, \eta_{sd}\}/\lambda_{sp}$.

**Theorem 5.** *A closed-form expression for the average achievable rate for symbol $s_1$ using MRC is given by*

$$\bar{C}_{s_1,MRC} = 0.5 \log_2(e) \left[ \frac{1}{\Gamma(N_r)\Omega_{sr}^{N_r}} \sum_{\nu=0}^{N_d-1} \frac{\left\{ G_{3,3}^{2,3}\left(\frac{Q}{\phi\Omega_{sp}} \middle| \begin{smallmatrix} 1-N_r-\nu, & 1, & 1 \\ 1, & 1, & 0 \end{smallmatrix}\right) - G_{3,3}^{2,3}\left(\frac{a_2 Q}{\phi\Omega_{sp}} \middle| \begin{smallmatrix} 1-N_r-\nu, & 1, & 1 \\ 1, & 1, & 0 \end{smallmatrix}\right) \right\}}{\Gamma(\nu+1)\Omega_{sd}^\nu \phi^{N_r+\nu}} \right.$$
$$\left. + \frac{1}{\Gamma(N_d)\Omega_{sd}^{N_d}} \sum_{\mu=0}^{N_r-1} \frac{\left\{ G_{3,3}^{2,3}\left(\frac{Q}{\phi\Omega_{sp}} \middle| \begin{smallmatrix} 1-N_d-\mu, & 1, & 1 \\ 1, & 1, & 0 \end{smallmatrix}\right) - G_{3,3}^{2,3}\left(\frac{a_2 Q}{\phi\Omega_{sp}} \middle| \begin{smallmatrix} 1-N_d-\mu, & 1, & 1 \\ 1, & 1, & 0 \end{smallmatrix}\right) \right\}}{\Gamma(\mu+1)\Omega_{sr}^\mu \phi^{N_d+\mu}} \right],$$
$$(25)$$

*where $\Gamma(\cdot)$ is the Gamma function and $G_{p,q}^{m,n}(\cdot \mid \cdot)$ is Meijer's G function.*

*Proof*: See Appendix E.

Similarly, the average achievable rate for symbol $s_2$ using MRC is given by

$$\bar{C}_{s_2,MRC} = 0.5 \int_0^\infty \log_2(1 + Qx) f_{\mathcal{Y}}(x)\, dx, \qquad (26)$$

where $\mathcal{Y} \triangleq \min\{\eta_{sr}a_2/\lambda_{sp}, \eta_{rd}/\lambda_{rp}\}$.

**Theorem 6.** *A closed-form expression for the average achievable rate for symbol $s_2$ using MRC is given by*

$$
\bar{C}_{s_2,\mathrm{MRC}} = \frac{0.5\log_2(e)N_r}{\Gamma(N_r+1)} G_{3,3}^{2,3}\left(\frac{a_2\Omega_{sr}Q}{\Omega_{sp}} \,\middle|\, \begin{matrix} 1-N_r,\,1,\,1 \\ 1,\quad 1,\,0 \end{matrix}\right) + \frac{0.5\log_2(e)N_d}{\Gamma(N_d+1)} G_{3,3}^{2,3}\left(\frac{\Omega_{rd}Q}{\Omega_{rp}} \,\middle|\, \begin{matrix} 1-N_d,\,1,\,1 \\ 1,\quad 1,\,0 \end{matrix}\right)
$$
$$
- \frac{0.5\log_2(e)}{\Gamma(N_r)\Gamma(N_d)}\left(\frac{a_2\Omega_{rp}\Omega_{sr}}{\Omega_{rd}\Omega_{sp}}\right)^{N_d} G_{1,\,1:2,\,2:2,\,2}^{1,1:1,2:1,2}\left(\begin{matrix} 1-N_r-N_d \\ 1-N_d \end{matrix} \,\middle|\, \begin{matrix} 1,\,1 \\ 1,\,0 \end{matrix} \,\middle|\, \begin{matrix} -N_d,\,1-N_d \\ 0,\quad -N_d \end{matrix} \,\middle|\, \frac{Qa_2\Omega_{sr}}{\Omega_{sp}},\,\frac{a_2\Omega_{sr}\Omega_{rp}}{\Omega_{sp}\Omega_{rd}}\right)
$$
$$
- \frac{0.5\log_2(e)}{\Gamma(N_d)\Gamma(N_r)}\left(\frac{\Omega_{sp}\Omega_{rd}}{a_2\Omega_{sr}\Omega_{rp}}\right)^{N_r} G_{1,\,1:2,\,2:2,\,2}^{1,1:1,2:1,2}\left(\begin{matrix} 1-N_r-N_d \\ 1-N_r \end{matrix} \,\middle|\, \begin{matrix} 1,\,1 \\ 1,\,0 \end{matrix} \,\middle|\, \begin{matrix} -N_r,\,1-N_r \\ 0,\quad -N_r \end{matrix} \,\middle|\, \frac{Q\Omega_{rd}}{\Omega_{rp}},\,\frac{\Omega_{rd}\Omega_{sp}}{a_2\Omega_{sr}\Omega_{rp}}\right),
$$
$$(27)$$

*where $G_{p_1,q_1:p_2,q_2:p_3,q_3}^{m_1,n_1:m_2,n_2:m_3,n_3}(\cdot\,|\,\cdot\,|\,\cdot\,|\,\cdot,\cdot)$ denotes the EGBMGF[2] [25].*

*Proof*: See Appendix F.

Using (25) and (27), the average achievable sum-rate for the SS-based CRS-NOMA using MRC is given by

$$
\bar{C}_{\mathrm{sum,MRC}} = \bar{C}_{s_1,\mathrm{MRC}} + \bar{C}_{s_2,\mathrm{MRC}}. \tag{28}
$$

For the case of CRS-OMA with MRC, the average achievable rate is given as

$$
\bar{C}_{\mathrm{OMA-MRC}} = 0.5\mathbb{E}_{\mathcal{Z}}\left[\log_2(1+Q\mathcal{Z})\right], \tag{29}
$$

where $\mathcal{Z} \triangleq \min\left\{\frac{\eta_{sr}}{\lambda_{sp}},\frac{\eta_{sd}}{\lambda_{sp}}+\frac{\eta_{rd}}{\lambda_{rp}}\right\}$.

*Outage probability for CRS-NOMA with MRC:*

Similar to the previous cases, we define $\mathcal{O}_1$ as the outage event for symbol $s_1$ in SS-based CRS-NOMA using MRC. Hence the outage probability for symbol $s_1$ is given by

$$
\mathrm{Pr}(\mathcal{O}_1) = \mathrm{Pr}(C_{s_1,\mathrm{MRC}} < R_1) = \mathrm{Pr}(\mathcal{X} < \Theta_1) = \int_0^{\Theta_1} f_{\mathcal{X}}(x)dx.
$$

---

[2]Efficient Mathematica® implementations of EGBMGF are given in [22, Table II] and [23]; and a Matlab® implementation is given in [24].

Substituting the expression for $f_{\mathcal{X}}(x)$ from (51) into the equation above and solving the integral using [20, Eqn. (3.194-1), p. 315], the outage probability for symbol $s_1$ in CRS-NOMA using MRC is given by

$$
\Pr(\mathcal{O}_1) = \underbrace{\frac{1}{\Gamma(N_r)\Omega_{sr}^{N_r}} \sum_{\nu=0}^{N_d-1} (\nu+1)_{N_r} \frac{(\Omega_{sp}\Theta_1)^{N_r+\nu}}{\Omega_{sd}^{\nu}(N_r+\nu)} \, {}_2F_1\left(N_r+\nu+1, N_r+\nu; N_r+\nu+1; -\Omega_{sp}\phi\Theta_1\right)}_{\mathcal{T}_1}
$$

$$
+ \underbrace{\frac{1}{\Gamma(N_d)\Omega_{sd}^{N_d}} \sum_{\mu=0}^{N_r-1} (\mu+1)_{N_d} \frac{(\Omega_{sp}\Theta_1)^{N_d+\mu}}{\Omega_{sr}^{\mu}(N_d+\mu)} \, {}_2F_1\left(N_d+\mu+1, N_d+\mu; N_d+\mu+1; -\Omega_{sp}\phi\Theta_1\right)}_{\mathcal{T}_2},
$$

$$\tag{30}$$

where $(x)_n \triangleq \Gamma(x+n)/\Gamma(x)$ denotes the Pochhammer symbol. Next, we define $\mathcal{O}_2$ as the outage event for symbol $s_2$ using MRC, similar to the previous cases. Hence, the outage probability for symbol $s_2$ is given by

$$
\Pr(\mathcal{O}_2) = F_{\frac{\eta_{sr}}{\lambda_{sp}}}(\Theta) + F_{\frac{\eta_{rd}}{\lambda_{rp}}}\left(\frac{\epsilon_2}{Q}\right) - F_{\frac{\eta_{sr}}{\lambda_{sp}}}(\Theta) F_{\frac{\eta_{rd}}{\lambda_{rp}}}\left(\frac{\epsilon_2}{Q}\right).
$$

Using (54), the analytical expression for the outage probability of symbol $s_2$ for SS-based CRS-NOMA using MRC is represented by

$$
\Pr(\mathcal{O}_2)
$$
$$
= \left(\frac{\Theta\Omega_{sp}}{\Omega_{sr}}\right)^{N_r} {}_2F_1\left(N_r+1, N_r; N_r+1; \frac{-\Theta\Omega_{sp}}{\Omega_{sr}}\right) + \left(\frac{\epsilon_2\Omega_{rp}}{Q\Omega_{rd}}\right)^{N_d} {}_2F_1\left(N_d+1, N_d; N_d+1; \frac{-\epsilon_2\Omega_{rp}}{Q\Omega_{rd}}\right)
$$
$$
- \left(\frac{\Theta\Omega_{sp}}{\Omega_{sr}}\right)^{N_r} \left(\frac{\epsilon_2\Omega_{rp}}{Q\Omega_{rd}}\right)^{N_d} {}_2F_1\left(N_r+1, N_r; N_r+1; \frac{-\Theta\Omega_{sp}}{\Omega_{sr}}\right) {}_2F_1\left(N_d+1, N_d; N_d+1; \frac{-\epsilon_2\Omega_{rp}}{Q\Omega_{sr}}\right).
$$

$$\tag{31}$$

On the other hand, the outage probability for the SS-based CRS-OMA with MRC is given by

$$
\Pr(\mathcal{O}_{OMA}) = \Pr(\mathcal{Z} < \epsilon_1).
$$

*Asymptotic behavior of CRS-NOMA with MRC*

Using the series expansion of the Gauss hypergeometric function [26, Eqn. (15.2.1). p. 384],

$$
\mathcal{T}_1 = \frac{1}{\Gamma(N_r)\Omega_{sr}^{N_r}} \sum_{\nu=0}^{N_d-1} \sum_{n=0}^{\infty} \frac{(\nu+1)_{N_r}\Omega_{sp}^{N_r+\nu}\epsilon_1^{N_r+\nu}}{\Omega_{sd}^{\nu}(N_r+\nu)(a_1-\epsilon_1 a_2)^{N_r+\nu}Q^{N_r+\nu}} \frac{(-1)^n(N_r+\nu)_n(\Omega_{sp}\phi\epsilon_1)^n}{n!(a_1-\epsilon_1 a_2)^n Q^n}
$$
$$
= \left[\frac{\Omega_{sp}\epsilon_1}{\Omega_{sr}(a_1-\epsilon_1 a_2)Q}\right]^{N_r} + \mathbf{O}\left(Q^{-(N_r+1)}\right).
$$

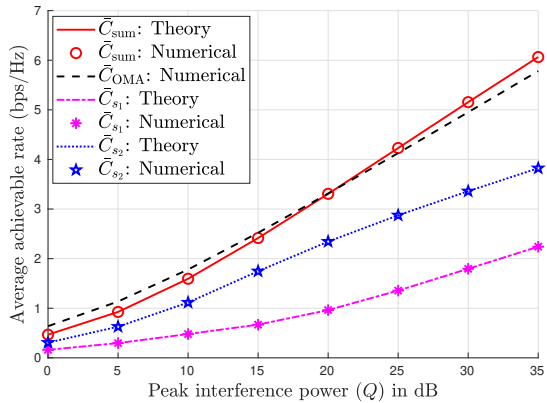

Fig. 2: Comparison of the average achievable rate for the SS-based CRS with $N_r = N_d = 1$.

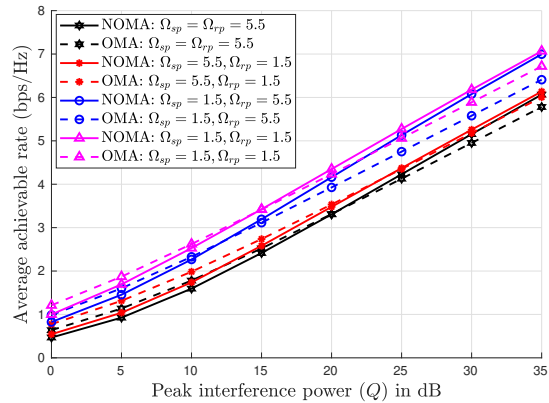

Fig. 3: Effect of the interference channels on the average achievable rate for the SS-based CRS.

Therefore, it is clear that $\mathcal{T}_1$ decays as $Q^{-N_r}$ for large values of $Q$. Using similar arguments, it can be shown that $\mathcal{T}_2$ in (30) decays as $Q^{-N_d}$ for large values of $Q$. Hence, it is straightforward to conclude that the outage probability for $s_1$ in CRS-NOMA with MRC decays as $Q^{-\min\{N_r, N_d\}}$ for large values of $Q$. Similarly, using the series expansion of the Gauss hypergeometric functions in (31), it can be shown that the outage probability for $s_2$ in CRS-NOMA with MRC decays as $Q^{-\min\{N_r, N_d\}}$ for large values of $Q$.

## VI. Results and Discussion

In this section, we present the analytical and numerical[3] results for the average achievable rate and outage probability for the spectrum sharing based cooperative relaying system. We consider the SS-based CRS system where $\Omega_{sd} = 1$, $\Omega_{sr} = \Omega_{rd} = 10$ and $\Omega_{sp} = \Omega_{rp} = 5.5$ (unless otherwise stated). For all NOMA based systems, we consider $R_1 = R_2 = 1$ bps/Hz.

The necessary constraint $a_1 > \epsilon_1 a_2$ (as noted in Section III) implies a possible range $0 < a_2 < 2^{-2R_1}$. Therefore the optimization of $a_2$ was performed using a one-dimensional search over the M-element discrete set $a_2 \in \{\epsilon, 2\epsilon, 3\epsilon, \ldots, M\epsilon\}$, where M is a positive integer and $\epsilon = 2^{-2R_1}/(M+1)$. In this paper, we consider $M = 24$ ($\epsilon = 0.01$). All results for NOMA-based systems presented in this section will be for the optimized power allocation.

Fig. 2 shows a comparison of the average achievable rate for the SS-based CRS with $N_r = N_d = 1$. It is clear from the figure that for small values of peak interference power constraint $Q$, the SS-based CRS-OMA gives higher achievable rate, but for large values of Q, the SS-based CRS-NOMA outperforms its OMA-based counterpart and achieves higher spectral efficiency. This is due to the fact that for small values of Q, the average transmit power from the SU-Tx and the relay is small, and for small transmit SNR, the average achievable sum-rate of CRS-NOMA is less than the average achievable rate of CRS-OMA as shown in [10, Fig. 1].

---

[3]For our numerical results, we do not realize the actual scenario, but rather generate the random variables and then evaluate the average achievable rate.

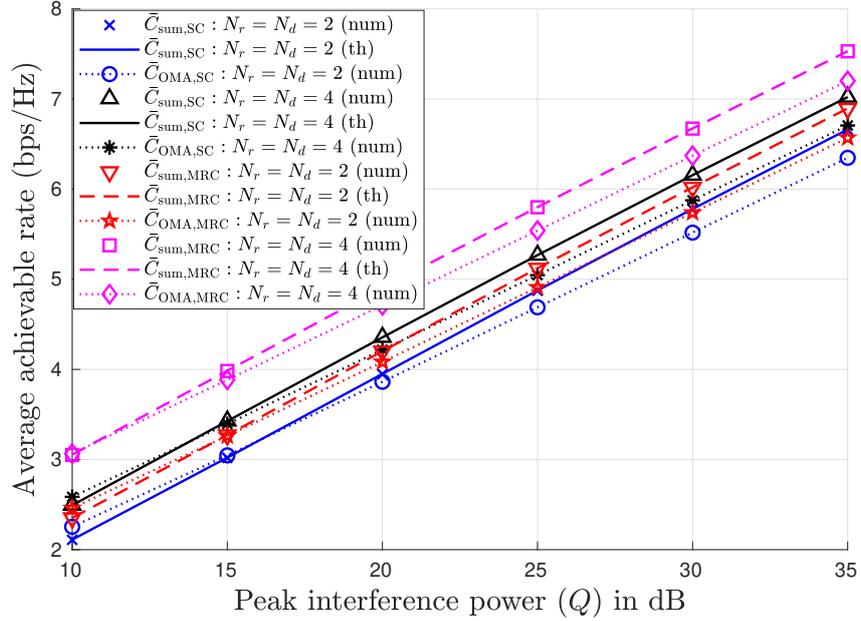

Fig. 4: Comparison of average achievable rate for SS-based CRS using SC and MRC for $\Omega_{\mathrm{sp}} = \Omega_{\mathrm{sp}} = 5.5$.

Fig. 3 shows the effect of the mean-square value of the interference links (i.e., $\Omega_{\mathrm{sp}}$ and $\Omega_{\mathrm{rp}}$) on the average achievable rate. The figures shows that when the interference links are weaker (in terms of mean-square value) the CRS achieves higher spectral efficiency. Moreover, it is clear from the figure that when the S-P link is weaker as compared to the R-P link (i.e., $\Omega_{\mathrm{sp}} = 1.5$ and $\Omega_{\mathrm{rp}} = 5.5$), the achievable rate of the CRS is comparatively higher than for the case when the former is stronger (i.e., $\Omega_{\mathrm{sp}} = 5.5$ and $\Omega_{\mathrm{rp}} = 1.5$). This means that the S-P interference channel has a more severe effect on the overall achievable rate of the CRS. This is due to the fact that when the interference channel between the SU-Tx and the PU-Rx is strong (i.e., $\Omega_{\mathrm{sp}}$ is large), the average power transmitted from the SU-Tx is small which adversely effects the achievable rate of both $s_1$ and $s_2$, while for the case when the interference channel between the relay and the PU-Rx is strong (i.e., $\Omega_{\mathrm{rp}}$ is large) the average power transmitted from the relay

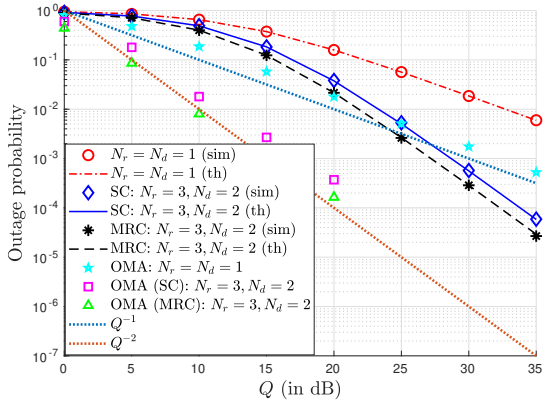

Fig. 5: Outage probability of symbol $s_1$ for the SS-based CRS.

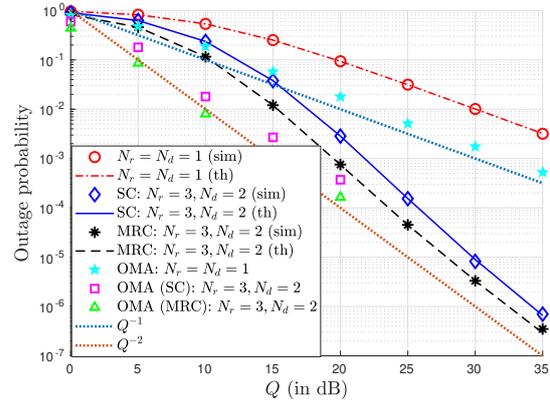

Fig. 6: Outage probability of symbol $s_2$ for the SS-based CRS.

is small which affects the achievable rate of symbol $s_2$ only.

Fig. 4 shows the average achievable rate for the SS-based CRS using selection combining and maximal-ratio combining for different values of $N_r$ and $N_d$. It is evident from the figure that with an increase in the number of antennas at the relay and at the SU-Rx, the average achievable rate increases for both NOMA and OMA systems. Also, it is important to note that with this increase in the number of antennas, the threshold value of $Q$ at which the CRS-NOMA outperforms its OMA-based counterpart becomes lower. Moreover, it is evident from the figure that CRS with MRC achieves significantly higher spectral efficiency as compared to that of CRS with SC for both NOMA and OMA systems.

Figs. 5 and 6 show the outage probabilities for SS-based CRS for $s_1$ and $s_2$, respectively (in the case of SS-based CRS-OMA, both these figures show the outage probability of $s_1$ with a target data rate of 1 bps/Hz). It is evident from the figures that the outage probability for SS-based CRS-NOMA decreases significantly with an increase in the number of antennas at the relay and the SU-Rx. Also, it can be noted from the figures that the outage probability of both the symbols with MRC is significantly lower than that with SC. These figures also confirm that for large values of $Q$, the outage probabilities decay as $Q^{-\min\{N_r, N_d\}}$, as proved analytically in the previous sections. The outage probability for the NOMA-based system is worse than that of the OMA-based system, because of the existence of co-channel interference and a higher target data rate requirement in the case of CRS-NOMA[4]. Also it is important to note that the outage probability of symbol $s_1$ is higher than that of symbol $s_2$ for SS-based CRS-NOMA. This happens because the outage probability of $s_1$ is dominated by the links between SU-Tx and SU-Rx, which are the weakest (on average).

## VII. Conclusion

In this paper, we provided a comprehensive average achievable sum-rate and outage probability analysis of a NOMA-based cooperative relaying system with underlay spectrum sharing considering a peak interference constraint in Rayleigh fading. We considered scenarios where the relay and the secondary receiver are equipped with multiple receive antennas and where both apply selection combining/maximal-ratio combining for signal reception. We also derived simplified analytical expressions for the special case when there is a single receive antenna at both relay and secondary-user receiver. Our results show that the optimal power allocation for maximizing the achievable sum-rate of CRS-NOMA depends on the target data rate requirement. It was shown that for higher values of peak interference power $Q$, the spectrum sharing system based on CRS-NOMA outperforms the spectrum sharing system based on conventional CRS-OMA, achieving higher spectral efficiency, for both SC and MRC schemes. Results also indicate that the interference channel between the secondary-user transmitter and the primary-user receiver has a more severe effect on the spectral efficiency of the secondary network as compared to that between the relay to the primary-user receiver interference channel. For a better insight into system behavior, we also presented the asymptotic (in the peak interference power) outage behavior for the SS-based CRS-NOMA. Our results also indicate that significant capacity gains can be achieved when the interference channel between the secondary-user transmitter/relay to primary-user receiver is in a deep fade.

## Appendix A
## Proof of Theorem 1

Since $|h_i|, i \in \{sr, sd, rd, sp, rp\}$ is Rayleigh distributed, the PDF and the CDF of $\lambda_i = |h_i|^2$ are, respectively, given by

$$f_{\lambda_i}(x) = \frac{1}{\Omega_i} \exp\left(\frac{-x}{\Omega_i}\right), \ F_{\lambda_i}(x) = 1 - \exp\left(\frac{-x}{\Omega_i}\right). \qquad (32)$$

---

[4]Note that in the case of CRS-NOMA, *two symbols* were transmitted in the first time slot with target data rate of 1 bps/Hz for each, whereas in the case of CRS-OMA, only a *single symbol* was transmitted in the first time slot with a target data rate of 1 bps/Hz.

Therefore, the PDF of $\min\{\lambda_{sr}, \lambda_{sd}\}$ is given by[5]

$$f_{\min\{\lambda_{sr}, \lambda_{sd}\}}(x) = f_{\lambda_{sr}}(x)[1 - F_{\lambda_{sd}}(x)] + f_{\lambda_{sd}}(x)[1 - F_{\lambda_{sr}}(x)] = \phi\exp(-\phi x),$$

where $\phi = (1/\Omega_{sr}) + (1/\Omega_{sd})$. The PDF of $X = \min\{\lambda_{sr}, \lambda_{sd}\}/\lambda_{sp}$ is therefore given by

$$f_X(x) = \int_0^\infty y f_{\min\{\lambda_{sr}, \lambda_{sd}\}}(yx) f_{\lambda_{sp}}(y), dy = \frac{\phi}{\Omega_{sp}} \int_0^\infty y \exp\left[-\left(\phi x + \frac{1}{\Omega_{sp}}\right)y\right] dy = \frac{\phi\Omega_{sp}}{(1 + \phi\Omega_{sp}x)^2}. \tag{33}$$

The integral above is solved using [20, Eqn. (3.351-3), p. 340]. Using (33), the first integral in (3) can be solved as

$$I_1 \triangleq \log_2(e)\phi\Omega_{sp} \int_0^\infty \frac{\ln(1 + Qx)dx}{(1 + \phi\Omega_{sp}x)^2} = \frac{Q\log_2\left(\frac{Q}{\phi\Omega_{sp}}\right)}{Q - \phi\Omega_{sp}}. \tag{34}$$

The integration above is solved using [20, Eqn. (4.291-17)]. Similarly, the second integral in (3) can be solved by replacing $Q$ in (34) with $a_2 Q$. Hence, using (3) and (34), the closed-form expression for the average achievable rate for symbol $s_1$ in SS-based CRS-NOMA system reduces to (4).

## Appendix B
## Proof of Theorem 2

Using (32) and a transformation of random variables, we have

$$f_{\lambda_{sr}a_2}(x) = \frac{1}{a_2} f_{\lambda_{sr}}\left(\frac{x}{a_2}\right) = \frac{1}{a_2\Omega_{sr}}\exp\left(\frac{-x}{a_2\Omega_{sr}}\right).$$

Hence,

$$f_{\lambda_{sr}a_2/\lambda_{sp}}(x) = \int_0^\infty y f_{\lambda_{sr}a_2}(yx) f_{\lambda_{sp}}(y)\, dy = \frac{1}{a_2\Omega_{sr}\Omega_{sp}} \int_0^\infty y \exp\left[-\left(\frac{x}{a_2\Omega_{sr}} + \frac{1}{\Omega_{sp}}\right)y\right] dy$$

$$= \frac{a_2\Omega_{sr}\Omega_{sp}}{(a_2\Omega_{sr} + \Omega_{sp}x)^2}, \qquad\qquad\qquad (\text{using } [20, 3.351\text{-}3, \text{ p. } 340])$$

and

$$F_{\lambda_{sr}a_2/\lambda_{sp}}(x) = \int_0^x f_{\lambda_{sr}a_2/\lambda_{sp}}(t)\, dt = \frac{\Omega_{sp}x}{a_2\Omega_{sr} + \Omega_{sp}x}. \tag{35}$$

---

[5]Given two random variables $\mathcal{U}$ and $\mathcal{V}$ with PDFs $f_\mathcal{U}(x)$ and $f_\mathcal{V}(x)$ respectively, and CDFs $F_\mathcal{U}(x)$ and $F_\mathcal{V}(x)$ respectively, the PDF of $\mathcal{W} \triangleq \min\{\mathcal{U}, \mathcal{V}\}$ is given by $f_\mathcal{W}(x) = f_\mathcal{U}(x)[1 - F_\mathcal{V}(x)] + f_\mathcal{V}[1 - F_\mathcal{U}(x)]$ and the CDF of $\mathcal{W}$ is given by $F_\mathcal{W}(x) = F_\mathcal{U}(x) + F_\mathcal{V}(x) - F_\mathcal{U}(x)F_\mathcal{V}(x)$.

The integration above is solved using [20, Eqn. (3.194-1)]. Similarly, the CDF of $\lambda_{rd}/\lambda_{rp}$ can be expressed by replacing $a_2\Omega_{sr}$ and $\Omega_{sp}$ in (35) with $\Omega_{rd}$ and $\Omega_{rp}$, respectively. Therefore, we have

$$1 - F_Y(x) = 1 - F_{\frac{\lambda_{sr}a_2}{\lambda_{sp}}}(x) - F_{\frac{\lambda_{rd}}{\lambda_{rp}}}(x) + F_{\frac{\lambda_{rd}}{\lambda_{rp}}}(x)F_{\frac{\lambda_{sr}a_2}{\lambda_{sp}}}(x) = \frac{a_2\Omega_{sr}\Omega_{rd}}{(a_2\Omega_{sr} + \Omega_{sp}x)(\Omega_{rd} + \Omega_{rp}x)}. \tag{36}$$

Using (8) and (36), the average achievable rate for symbol $s_2$ is given as

$$\bar{C}_{s_2} = \lim_{\Lambda \to \infty} \int_0^\Lambda \frac{0.5\log_2(e)\,Q\,a_2\,\Omega_{sr}\Omega_{rd}\,dx}{(a_2\Omega_{sr} + \Omega_{sp}x)(\Omega_{rd} + \Omega_{rp}x)(1 + Qx)}. \tag{37}$$

Solving the integral above using partial fractions, (37) reduces to (9); this completes the proof.

## Appendix C
## Proof of Theorem 3

The CDF and PDF of $\delta_{sr}$ (maximum of exponentially-distributed independent random variables) are, respectively,

$$F_{\delta_{sr}}(x) = 1 - \sum_{k=1}^{N_r}(-1)^{k-1}\binom{N_r}{k}\exp\left(\frac{-kx}{\Omega_{sr}}\right), \tag{38}$$

$$f_{\delta_{sr}}(x) = \sum_{k=1}^{N_r}(-1)^{k-1}\binom{N_r}{k}\frac{k}{\Omega_{sr}}\exp\left(\frac{-kx}{\Omega_{sr}}\right). \tag{39}$$

The CDF and PDF of $\delta_{sd}$ can be represented by replacing $N_r$ and $\Omega_{sr}$, respectively, by $N_d$ and $\Omega_{sd}$, respectively, in both (38) and (39). The PDF of $\min\{\delta_{sr}, \delta_{sd}\}$ is given by

$$f_{\min\{\delta_{sr},\delta_{sd}\}}(x) = \sum_{k=1}^{N_r}\sum_{j=1}^{N_d}(-1)^{k+j}\binom{N_r}{k}\binom{N_d}{j}\xi_{k,j}\exp(-\xi_{k,j}x),$$

where $\xi_{k,j} = (k/\Omega_{sr}) + (j/\Omega_{sd})$. The PDF of $\mathcal{X} = \frac{\min\{\delta_{sr},\delta_{sd}\}}{\lambda_{sp}}$ is therefore given by

$$f_{\mathcal{X}}(x) = \int_0^\infty y f_{\min\{\delta_{sr},\delta_{sd}\}}(yx)f_{\lambda_{sp}}(y)\,dy$$

$$= \sum_{k=1}^{N_r}\sum_{j=1}^{N_d}\binom{N_r}{k}\binom{N_d}{j}\frac{(-1)^{k+j}\xi_{k,j}}{\Omega_{sp}}\int_0^\infty y\exp\left[-\left(\xi_{k,j}x + \frac{1}{\Omega_{sp}}\right)y\right]dy$$

$$= \sum_{k=1}^{N_r}\sum_{j=1}^{N_d}\binom{N_r}{k}\binom{N_d}{j}\frac{(-1)^{k+j}\xi_{k,j}}{\Omega_{sp}}\left(\xi_{k,j}x + \frac{1}{\Omega_{sp}}\right)^{-2}. \tag{40}$$

The integral above is solved using [20, Eqn. (3.351-3), p. 340]. Now, the first integral in (15) can be solved as

$$
\begin{aligned}
I_2 &\triangleq \log_2(e) \int_0^\infty \ln(1+Qx) f_{\mathcal{X}}(x) dx \\
&= \sum_{k=1}^{N_r} \sum_{j=1}^{N_d} \binom{N_r}{k} \binom{N_d}{j} \frac{(-1)^{k+j} \xi_{k,j}}{\ln(2)\Omega_{sp}} \int_0^\infty \ln(1+Qx) \left(\xi_{k,j}x + \frac{1}{\Omega_{sp}}\right)^{-2} dx \\
&= \sum_{k=1}^{N_r} \sum_{j=1}^{N_d} \binom{N_r}{k} \binom{N_d}{j} \frac{(-1)^{k+j}Q}{Q - \xi_{k,j}\Omega_{sp}} \log_2\left(\frac{Q}{\xi_{k,j}\Omega_{sp}}\right).
\end{aligned} \tag{41}
$$

The integration above is solved using [20, Eqn. (4.291-17), p. 556]. Similarly, the second integral in (15) can be solved by replacing $Q$ in (41) by $a_2 Q$. Using (15) and (41), the closed-form expression for the average achievable rate for symbol $s_1$ in the CRS-NOMA system with SC reduces to (16); this completes the proof.

<div align="center">

Appendix D

Proof of Theorem 4

</div>

Using (39) and a transformation of random variables, the PDF of $\delta_{sr}a_2$ is given by

$$
f_{\delta_{sr}a_2}(x) = \frac{f_{\delta_{sr}}\left(\frac{x}{a_2}\right)}{a_2} = \sum_{k=1}^{N_r} \binom{N_r}{k} \frac{(-1)^{k-1}k}{a_2\Omega_{sr}} \exp\left(\frac{-kx}{a_2\Omega_{sr}}\right).
$$

The PDF of $\delta_{sr}a_2/\lambda_{sp}$ is given by

$$
\begin{aligned}
&f_{\delta_{sr}a_2/\lambda_{sp}}(x) = \int_0^\infty y f_{\delta_{sr}a_2}(yx) f_{\lambda_{sp}}(y) \, dy \\
&= \sum_{k=1}^{N_r} \binom{N_r}{k} \frac{(-1)^{k-1}k}{a_2\Omega_{sr}\Omega_{sp}} \int_0^\infty y \exp\left[-\left(\frac{kx}{a_2\Omega_{sr}} + \frac{1}{\Omega_{sp}}\right)y\right] dy = \sum_{k=1}^{N_r} \binom{N_r}{k} \frac{(-1)^{k-1}k}{a_2\Omega_{sr}\Omega_{sp}} \left(\frac{kx}{a_2\Omega_{sr}} + \frac{1}{\Omega_{sp}}\right)^{-2}.
\end{aligned}
$$

The integration above is solved using [20, Eqn. (3.351-3), p. 340]. Using [20, Eqn. (3.194-1), p. 315], the CDF of $\delta_{sr}a_2/\lambda_{sp}$ is given by

$$
F_{\frac{\delta_{sr}a_2}{\lambda_{sp}}}(x) = \int_0^x f_{\frac{\delta_{sr}a_2}{\lambda_{sp}}}(t) \, dt = \sum_{k=1}^{N_r} \binom{N_r}{k} \frac{(-1)^{k-1}k\Omega_{sp}x}{a_2\Omega_{sr} + k\Omega_{sp}x}. \tag{42}
$$

Similarly, the CDF of $\delta_{rd}/\delta_{rp}$ can be expressed by replacing $N_r$, $a_2\Omega_{sr}$ and $\Omega_{sp}$ in (42) by $N_d$, $\Omega_{rd}$ and $\Omega_{rp}$, respectively. Therefore, for $\mathcal{Y} = \min\left\{\frac{\delta_{sr}a_2}{\lambda_{sp}}, \frac{\delta_{rd}}{\lambda_{rp}}\right\}$, we have

$$
\begin{aligned}
1 - F_{\mathcal{Y}}(x) = 1 - F_{\frac{\delta_{sr}a_2}{\lambda_{sp}}}(x) - F_{\frac{\delta_{rd}}{\lambda_{rp}}} + F_{\frac{\delta_{sr}a_2}{\lambda_{sp}}}(x) F_{\frac{\delta_{rd}}{\lambda_{rp}}}(x) = 1 - \sum_{k=1}^{N_r} \binom{N_r}{k} \frac{(-1)^{k-1}k\Omega_{sp}x}{a_2\Omega_{sr} + k\Omega_{sp}x} \\
- \sum_{j=1}^{N_d} \binom{N_d}{j} \frac{(-1)^{j-1}j\Omega_{rp}x}{\Omega_{rd} + j\Omega_{rp}x} + \sum_{k=1}^{N_r} \sum_{j=1}^{N_d} \binom{N_r}{k} \binom{N_d}{j} \frac{(-1)^{k+j}kj\Omega_{sp}\Omega_{rp}x^2}{(a_2\Omega_{sr} + k\Omega_{sp}x)(\Omega_{rd} + j\Omega_{rp}x)}. \tag{43}
\end{aligned}
$$

Using (17), the average achievable rate for symbol $s_2$ using SC is given by

$$\bar{C}_{s_2,\mathrm{SC}} = \frac{1}{2}\int_0^\infty \log_2(1+Qx)f_{\mathcal{Y}}(x)dx = \frac{0.5Q}{\ln(2)}\int_0^\infty \frac{1-F_{\mathcal{Y}}(x)}{1+Qx}dx. \qquad (44)$$

Now we define the integral $I_3$ as

$$I_3 \triangleq \int_0^\infty \frac{1}{1+Qx}\left(\frac{k\Omega_{sp}x}{a_2\Omega_{sr}+k\Omega_{sp}x}\right)dx = \int_0^\infty \frac{1}{1+Qx}\left(1-\frac{a_2\Omega_{sr}}{a_2\Omega_{sr}+k\Omega_{sp}x}\right)dx$$

$$= \int_0^\infty \frac{1}{1+Qx}dx - \lim_{\Lambda\to\infty}\int_0^\Lambda \frac{a_2\Omega_{sr}\,dx}{(a_2\Omega_{sr}+k\Omega_{sp}x)(1+Qx)} = \int_0^\infty \frac{1}{1+Qx}dx - \frac{a_2\Omega_{sr}\ln\left(\frac{a_2\Omega_{sr}Q}{k\Omega_{sp}}\right)}{a_2\Omega_{sr}Q-k\Omega_{sp}}. \qquad (45)$$

The integration above is solved using partial fractions. Similarly,

$$I_4 \triangleq \int_0^\infty \frac{1}{1+Qx}\left(\frac{j\Omega_{rp}x}{\Omega_{rd}+j\Omega_{rp}x}\right)dx = \int_0^\infty \frac{1}{1+Qx}dx - \frac{\Omega_{rd}}{\Omega_{rd}Q-j\Omega_{rp}}\ln\left(\frac{\Omega_{rd}Q}{j\Omega_{rp}}\right), \qquad (46)$$

and

$$I_5 \triangleq \int_0^\infty \frac{1}{1+Qx}\left(\frac{kj\Omega_{sp}\Omega_{rp}x^2}{(a_2\Omega_{sr}+k\Omega_{sp}x)(\Omega_{rd}+j\Omega_{rp}x)}\right)dx$$

$$= \int_0^\infty \frac{1}{1+Qx}\left[1+\frac{k\Omega_{rd}^2\Omega_{sp}}{(a_2j\Omega_{rp}\Omega_{sr}-k\Omega_{rd}\Omega_{sp})(\Omega_{rd}+j\Omega_{rp}x)}+\frac{a_2^2j\Omega_{rp}\Omega_{sr}^2}{(k\Omega_{rd}\Omega_{sp}-a_2j\Omega_{rp}\Omega_{sr})(a_2\Omega_{sr}+k\Omega_{sp}x)}\right]dx$$

$$= \int_0^\infty \frac{1}{1+Qx}dx + \frac{k\Omega_{rd}^2\Omega_{sp}\ln\left(\frac{j\Omega_{rp}}{\Omega_{rd}Q}\right)}{(k\Omega_{rd}\Omega_{sp}-a_2j\Omega_{rp}\Omega_{sr})(\Omega_{rd}Q-j\Omega_{rp})}+\frac{a_2^2j\Omega_{rp}\Omega_{sr}^2\ln\left(\frac{a_2\Omega_{sr}Q}{k\Omega_{sp}}\right)}{(k\Omega_{rd}\Omega_{sp}-a_2j\Omega_{rp}\Omega_{sr})(a_2\Omega_{sr}Q-k\Omega_{sp})}. \qquad (47)$$

Moreover, we also have

$$\left[1-\sum_{k=1}^{N_r}(-1)^{k-1}\binom{N_r}{k}-\sum_{j=1}^{N_d}(-1)^{j-1}\binom{N_d}{j}+\sum_{k=1}^{N_r}\sum_{j=1}^{N-d}(-1)^{k+j}\binom{N_r}{k}\binom{N_d}{j}\right]\int_0^\infty \frac{dx}{1+Qx}=0. \qquad (48)$$

Using $(44)-(48)$, the closed-form expression for the average achievable rate of symbol $s_2$ using SC reduces to (18); this completes the proof.

<div align="center">

Appendix E

Proof of Theorem 5

</div>

Since $|h_{sr,i}|, \forall i \in \{1,2,\ldots,N_r\}$ is a Rayleigh distributed random variable, $\eta_{sr}$ will be Gamma distributed. Therefore,

$$f_{\eta_{sr}}(x) = \frac{x^{N_r-1}}{\Gamma(N_r)\Omega_{sr}^{N_r}}\exp\left(\frac{-x}{\Omega_{sr}}\right), \qquad (49)$$

$$F_{\eta_{sr}}(x) = 1-\exp\left(\frac{-x}{\Omega_{sr}}\right)\sum_{\mu=0}^{N_r-1}\frac{1}{\mu!}\left(\frac{x}{\Omega_{sr}}\right)^\mu. \qquad (50)$$

Similarly, the expression for $f_{\eta_{sd}}(x)$ and $F_{\eta_{sd}}(x)$ can be represented by replacing $N_r$ and $\Omega_{sr}$, respectively, by $N_d$ and $\Omega_{sd}$, respectively in both (49) and (50). Therefore,

$$
\begin{aligned}
f_{\min\{\eta_{sr},\eta_{sd}\}}(x) &= f_{\eta_{sr}}(x)[1 - F_{\eta_{sd}}(x)] + f_{\eta_{sd}}(x)[1 - F_{\eta_{sr}}(x)] \\
&= \frac{\exp(-\phi x)}{\Gamma(N_r)\Omega_{sr}^{N_r}} \sum_{\nu=0}^{N_d-1} \frac{x^{N_r+\nu-1}}{\nu!\Omega_{sd}^{\nu}} + \frac{\exp(-\phi x)}{\Gamma(N_d)\Omega_{sd}^{N_d}} \sum_{\mu=0}^{N_r-1} \frac{x^{N_d+\mu-1}}{\mu!\Omega_{sr}^{\mu}}.
\end{aligned}
$$

Using a transformation of random variables yields

$$
\begin{aligned}
f_{\chi}(x) &= \int_0^{\infty} y f_{\min\{\eta_{sr},\eta_{sd}\}}(yx) f_{\lambda_{sp}}(y) dy = \sum_{\nu=0}^{N_d-1} \frac{x^{N_r+\nu-1}}{\Gamma(N_r)\Omega_{sr}^{N_r}\Omega_{sp}\nu!\Omega_{sd}^{\nu}} \int_0^{\infty} y^{N_r+\nu} \exp\left[-\left(\phi x+\frac{1}{\Omega_{sp}}\right)y\right] dy \\
&+ \sum_{\mu=0}^{N_r-1} \frac{x^{N_d+\mu-1}}{\Gamma(N_d)\Omega_{sd}^{N_d}\Omega_{sp}\mu!\Omega_{sr}^{\mu}} \int_0^{\infty} y^{N_d+\mu} \exp\left[-\left(\phi x+\frac{1}{\Omega_{sp}}\right)y\right] dy = \sum_{\nu=0}^{N_d-1} \frac{x^{N_r+\nu-1}\Gamma(N_r+\nu+1)}{\Gamma(N_r)\Gamma(\nu+1)\Omega_{sr}^{N_r}\Omega_{sd}^{\nu}\Omega_{sp}} \\
&\times \left(\phi x+\frac{1}{\Omega_{sp}}\right)^{-(N_r+\nu+1)} + \sum_{\mu=0}^{N_r-1} \frac{x^{N_d+\mu-1}\Gamma(N_d+\mu+1)}{\Gamma(N_d)\Gamma(\mu+1)\Omega_{sd}^{N_d}\Omega_{sr}^{\mu}\Omega_{sp}} \left(\phi x+\frac{1}{\Omega_{sp}}\right)^{-(N_d+\mu+1)}.
\end{aligned}
$$
(51)

The integration above is solved using [20, Eqn. (3.351-3), p 340]. Substituting the expression for $f_{\chi}(x)$ from (51) into (24), and using [27, Eqn. (11)], the first integral in (24) can be solved as

$$
\begin{aligned}
I_6 &\triangleq 0.5\log_2(e) \int_0^{\infty} G_{2,2}^{1,2}\left(Qx \,\middle|\, \begin{matrix} 1, 1 \\ 1, 0 \end{matrix}\right) f_{\chi}(x) dx \\
&= 0.5\log_2(e) \left[ \sum_{\nu=0}^{N_d-1} \frac{\Gamma(N_r+\nu+1) \int_0^{\infty} x^{N_r+\nu-1}\left(x+\frac{1}{\Omega_{sp}\phi}\right)^{-(N_r+\nu+1)} G_{2,2}^{1,2}\left(Qx \,\middle|\, \begin{matrix} 1, 1 \\ 1, 0 \end{matrix}\right) dx}{\Gamma(N_r)\Gamma(\nu+1)\Omega_{sr}^{N_r}\Omega_{sd}^{\nu}\Omega_{sp}\phi^{N_r+\nu+1}} \right. \\
&\left. + \sum_{\mu=0}^{N_r-1} \frac{\Gamma(N_d+\mu+1) \int_0^{\infty} x^{N_d+\mu-1}\left(x+\frac{1}{\phi\Omega_{sp}}\right)^{-(N_d+\mu+1)} G_{2,2}^{1,2}\left(Qx \,\middle|\, \begin{matrix} 1, 1 \\ 1, 0 \end{matrix}\right) dx}{\Gamma(N_d)\Gamma(\mu+1)\Omega_{sd}^{N_d}\Omega_{sr}^{\mu}\Omega_{sp}\phi^{N_d+\mu+1}} \right] \\
&= 0.5\log_2(e) \left[ \sum_{\nu=0}^{N_d-1} \frac{G_{3,3}^{2,3}\left(\frac{Q}{\phi\Omega_{sp}} \,\middle|\, \begin{matrix} 1-N_r-\nu, 1, 1 \\ 1, \quad 1, 0 \end{matrix}\right)}{\Gamma(N_r)\Gamma(\nu+1)\Omega_{sr}^{N_r}\Omega_{sd}^{\nu}\phi^{N_r+\nu}} + \cdot \sum_{\mu=0}^{N_r-1} \frac{G_{3,3}^{2,3}\left(\frac{Q}{\phi\Omega_{sp}} \,\middle|\, \begin{matrix} 1-N_d-\mu, 1, 1 \\ 1, \quad 1, 0 \end{matrix}\right)}{\Gamma(N_d)\Gamma(\mu+1)\Omega_{sd}^{N_d}\Omega_{sr}^{\mu}\phi^{N_d+\mu}} \right].
\end{aligned}
$$
(52)

The integration above is solved using [20, Eqn. (7.811-5), p. 852]. Similarly, the second integral in (24) can be solved by replacing $Q$ in (52) with $a_2Q$. Therefore, using (24) and (52), the analytical expression for the average achievable rate of symbol $s_1$ in the SS-based CRS-NOMA using MRC reduces to (25).

## Appendix F

## Proof of Theorem 6

Using (49) and a transformation of random variables, we have

$$f_{\eta_{sr}a_2}(x) = \frac{1}{a_2}f_{\eta_{sr}}\left(\frac{x}{a_2}\right) = \frac{x^{N_r-1}}{\Gamma(N_r)\Omega_{sr}^{N_r}a_2^{N_r}}\exp\left(\frac{-x}{a_2\Omega_{sr}}\right).$$

Therefore,

$$f_{\eta_{sr}a_2/\lambda_{sp}}(x) = \int_0^\infty y f_{\eta_{sr}a_2}(yx)f_{\lambda_{sp}}(y)dy$$

$$= \frac{x^{N_r-1}}{\Gamma(N_r)\Omega_{sr}^{N_r}a_2^{N_r}\Omega_{sp}}\int_0^\infty y^{N_r}\exp\left[-\left(\frac{x}{a_2\Omega_{sr}}+\frac{1}{\Omega_{sp}}\right)y\right]dy = \frac{N_r\Omega_{sr}a_2\Omega_{sp}^{N_r}x^{N_r-1}}{(a_2\Omega_{sr}+\Omega_{sp}x)^{N_r+1}}. \tag{53}$$

The integration above is solved using [20, Eqn. (3.351-3), p.340]. Using (53), we have

$$F_{\frac{\eta_{sr}a_2}{\lambda_{sp}}}(x) = N_r\Omega_{sr}a_2\Omega_{sp}^{N_r}\int_0^x \frac{t^{N_r-1}dt}{(a_2\Omega_{sr}+\Omega_{sp}t)^{N_r+1}} = \left(\frac{\Omega_{sp}x}{a_2\Omega_{sr}}\right)^{N_r}{}_2F_1\left(N_r+1,N_r;N_r+1;\frac{-\Omega_{sp}}{a_2\Omega_{sr}}x\right), \tag{54}$$

where ${}_2F_1(\cdot,\cdot;\cdot;\cdot)$ is the Gauss hypergeometric function and the integration above is solved using [20, Eqn. (3.194-1), p. 315]. The PDF of $\eta_{rd}/\lambda_{rp}$ can be expressed by replacing $N_r, a_2\Omega_{sr}$ and $\Omega_{sp}$, respectively, in (53) by $N_d, \Omega_{rd}$ and $\Omega_{rp}$, respectively. The CDF of $\eta_{rd}/\lambda_{rp}$ can be expressed in a similar fashion using (54). Therefore, the PDF of $\mathcal{Y}$ can be obtained as

$$f_{\mathcal{Y}}(x) = f_{\eta_{sr}a_2/\lambda_{sp}}(x)[1-F_{\eta_{rd}/\lambda_{rp}}(x)] + f_{\eta_{rd}/\lambda_{rp}}(x)[1-F_{\eta_{sr}a_2/\lambda_{sp}}(x)] = \frac{N_r\Omega_{sr}a_2\Omega_{sp}^{N_r}x^{N_r-1}}{(a_2\Omega_{sr}+\Omega_{sp}x)^{N_r+1}}$$

$$+ \frac{N_d\Omega_{rd}\Omega_{sp}^{N_d}x^{N_d-1}}{(\Omega_{rd}+\Omega_{rp}x)^{N_d+1}} - \frac{N_r\Omega_{sr}a_2\Omega_{sp}^{N_r}\Omega_{rp}^{N_d}x^{N_r+N_d-1}}{\Omega_{rd}^{N_d}(a_2\Omega_{sr}+\Omega_{sp}x)^{N_r+1}}{}_2F_1\left(N_d+1,N_d;N_d+1;\frac{-\Omega_{rp}}{\Omega_{rd}}x\right)$$

$$- \frac{N_d\Omega_{rd}\Omega_{rp}^{N_d}\Omega_{sp}^{N_r}x^{N_r+N_d-1}}{a_2^{N_r}\Omega_{sr}^{N_r}(\Omega_{rd}+\Omega_{rp}x)^{N_d+1}}{}_2F_1\left(N_r+1,N_r;N_r+1;\frac{-\Omega_{sp}}{a_2\Omega_{sr}}x\right). \tag{55}$$

Using (26) and (55), we have

$$\bar{C}_{s_2,MRC} = 0.5\log_2(e)\left[I_7+I_8-I_9-I_{10}\right], \tag{56}$$

where

$$I_7 \triangleq \int_0^\infty \ln(1+Qx)\frac{N_r\Omega_{sr}a_2\Omega_{sp}^{N_r}x^{N_r-1}}{(a_2\Omega_{sr}+\Omega_{sp}x)^{N_r+1}}dx$$

$$= \frac{N_r\Omega_{sr}a_2}{\Omega_{sp}}\int_0^\infty G_{2,2}^{1,2}\left(Qx\,\Big|\,\begin{matrix}1,1\\1,0\end{matrix}\right)x^{N_r-1}\left(x+\frac{a_2\Omega_{sr}}{\Omega_{sp}}\right)^{-(N_r+1)}dx = \frac{N_r}{\Gamma(N_r+1)}G_{3,3}^{2,3}\left(\frac{a_2\Omega_{sr}Q}{\Omega_{sp}}\,\Big|\,\begin{matrix}1-N_r,\,1,\,1\\1,\quad1,\,0\end{matrix}\right). \tag{57}$$

The integration above is solved using [20, Eqn. (7.811-5), p. 852]. Similarly,

$$I_8 \triangleq \int_0^\infty \ln(1+Qx)\frac{N_d\Omega_{rd}\Omega_{sp}^{N_d}x^{N_d-1}}{(\Omega_{rd}+\Omega_{rp}x)^{N_d+1}}dx = \frac{N_d}{\Gamma(N_d+1)}G_{3,3}^{2,3}\left(\frac{\Omega_{rd}Q}{\Omega_{rp}}\,\Big|\,\begin{matrix}1-N_d,\,1,\,1\\1,\quad1,\,0\end{matrix}\right). \tag{58}$$

$$I_9 \triangleq \frac{N_r \Omega_{sr} a_2 \Omega_{sp}^{N_r} \Omega_{rp}^{N_d}}{\Omega_{rd}^{N_d}} \int_0^\infty \ln(1 + Qx) \frac{x^{N_r + N_d - 1}}{(a_2 \Omega_{sr} + \Omega_{sp} x)^{N_r + 1}} \, {}_2F_1 \left( N_d + 1, N_d; N_d + 1; \frac{-\Omega_{rp}}{\Omega_{rd}} x \right) dx.$$

Using [27, Eqns. (11), (17)] and [28], $I_9$ can be represented as

$$I_9 = \frac{N_r \Omega_{sp}^{N_r} \Omega_{rp}^{N_d} \int_0^\infty x^{N_r + N_d - 1} G_{2,2}^{1,2} \left( Qx \left| \begin{smallmatrix} 1, 1 \\ 1, 0 \end{smallmatrix} \right. \right) G_{1,1}^{1,1} \left( \frac{\Omega_{sp} x}{a_2 \Omega_{sr}} \left| \begin{smallmatrix} -N_r \\ 0 \end{smallmatrix} \right. \right) G_{2,2}^{1,2} \left( \frac{\Omega_{rp}}{\Omega_{rd}} x \left| \begin{smallmatrix} -N_d, \, 1-N_d \\ 0, \quad -N_d \end{smallmatrix} \right. \right) dx}{\Omega_{rd}^{N_d} a_2^{N_r} \Omega_{sr}^{N_r} \Gamma(N_r + 1) \Gamma(N_d)}$$

$$= \frac{1}{\Gamma(N_r) \Gamma(N_d)} \left( \frac{a_2 \Omega_{rp} \Omega_{sr}}{\Omega_{rd} \Omega_{sp}} \right)^{N_d} G_{1,1:2;2,2}^{1,1:1,2:1,2} \left( \begin{smallmatrix} 1-N_r-N_d \\ 1-N_d \end{smallmatrix} \left| \begin{smallmatrix} 1, 1 \\ 1, 0 \end{smallmatrix} \right| \begin{smallmatrix} -N_d, \, 1-N_d \\ 0, \quad -N_d \end{smallmatrix} \left| \frac{Q a_2 \Omega_{sr}}{\Omega_{sp}}, \frac{a_2 \Omega_{sr} \Omega_{rp}}{\Omega_{sp} \Omega_{rd}} \right. \right).$$

$$\tag{59}$$

The integration above is solved using [23, Eqn. (9)]. Similarly,

$$I_{10} = \frac{N_d \Omega_{rd} \Omega_{rp}^{N_d} \Omega_{sp}^{N_r}}{a_2^{N_r} \Omega_{sr}^{N_r}} \int_0^\infty \ln(1 + Qx) \frac{x^{N_r + N_d - 1}}{(\Omega_{rd} + \Omega_{rp} x)^{N_d + 1}} \, {}_2F_1 \left( N_r + 1, N_r; N_r + 1; \frac{-\Omega_{sp}}{a_2 \Omega_{sr}} x \right) dx$$

$$= \frac{1}{\Gamma(N_d) \Gamma(N_r)} \left( \frac{\Omega_{sp} \Omega_{rd}}{a_2 \Omega_{sr} \Omega_{rp}} \right)^{N_r} G_{1,1:2;2,2}^{1,1:1,2:1,2} \left( \begin{smallmatrix} 1-N_r-N_d \\ 1-N_r \end{smallmatrix} \left| \begin{smallmatrix} 1, 1 \\ 1, 0 \end{smallmatrix} \right| \begin{smallmatrix} -N_r, \, 1-N_r \\ 0, \quad -N_r \end{smallmatrix} \left| \frac{Q \Omega_{rd}}{\Omega_{rp}}, \frac{\Omega_{rd} \Omega_{sp}}{a_2 \Omega_{sr} \Omega_{rp}} \right. \right). \tag{60}$$

Using (26) and (56)-(60), the analytical expression for the average achievable rate of symbol $s_2$ in the SS-based CRS-NOMA using MRC reduces to (27); this completes the proof.